\documentclass{lmcs}

\usepackage{booktabs}   
\usepackage{subcaption} 

\usepackage{anyfontsize}
\usepackage[utf8]{inputenc}
\usepackage[T1]{fontenc} 
\usepackage{bm}
\usepackage{stackrel}

\usepackage{fancybox}
\usepackage[english]{babel}
\usepackage{mathpartir}
\usepackage{qtree}

\usepackage[export]{adjustbox}
\usepackage{caption} 
\usepackage{url}                  
\usepackage{listings}          
\usepackage{xcolor}
\definecolor{darkgreen}{rgb}{0,0.45,0.08} 

\usepackage{microtype}
\usepackage{balance}
\usepackage{tikz} 
\usepackage{pgf}
\usepackage{apxproof}   
\usepackage{multirow}
\usepackage{fancyvrb} 
\usepackage{version} 
\usepackage{relsize}
\usepackage{booktabs}

\let\phi\varphi
\newcommand{\temp}[1]{1}
\newcommand{\N}{\mathbb{N}}

\newcommand{\distinct}{\opname{distinct}}
\newcommand{\singleton}[2][]{\ensuremath{\ifstrempty{#1}{\bag{#2}}{\bag{#2}_#1} }}
\newcommand{\bag}[1]{\{\!\!\{ #1 \}\!\!\}}
\newcommand*{\setcomp}[2]{\ensuremath{\{#1 \mid  #2\}}}
\newcommand*{\bagcomp}[2]{\bag{#1\mid #2}}
\newcommand{\bunion}{\uplus}
\newcommand{\sunion}{\cup}
\def\emptybag{\bag{}}
\def\emptylist{\text{[ ]}}
\newcommand{\lunion}{\mathbin{++}}

\newcommand{\homo}[2][f]{\ensuremath{H_{#1}^{#2}}}
\newcommand{\mumonoid}{$\mu$-monoids}

\newcommand*\lambdaexp[2]{\ensuremath{\closure{#1}{#2}}}
\newcommand{\size}[1]{\ensuremath{\opname{size}(#1)}}
\newcommand{\sizetran}[1]{\ensuremath{\opname{size}_t(#1)}}
\renewcommand{\iff}{\Leftrightarrow}
\renewcommand{\implies}{\Rightarrow}

\newcommand\eqrel\sim
\newcommand\eqcomm{\eqrel_{\text{comm}}}
\newcommand\eqidem{\eqrel_{\text{idem}}}
\newcommand*\eqdelta[1][\delta]{\eqrel_{#1}}
\newcommand*\opdelta[1][\delta]{\otimes_{#1}}
\newcommand*\mondelta[1][\delta]{M_{#1}}

\newcommand*\opname[1]{\textsf{#1}}

\newcommand{\eqdef}{\stackrel{\text{\tiny def}}{=}}
\newcommand{\equivdef}{\stackrel{\text{\tiny def}}{\Leftrightarrow}}

\newcommand{\syntaxdef}{\mathrel{::=}}

\newcommand{\syntaxtable}[1]{
  \def\entry##1[##2]##3[##4]{
    {##1} & \syntaxdef & \hspace{3cm} & \!\!\!\! \mbox{##2}
    \\    &     & {##3} & \mbox{##4} }
  \def\singleentry##1[]##2[##3]{
  {##1} & \syntaxdef & {##2} & \!\!\!\! \mbox{##3} }
  \def\oris##1[##2]{
    \\    & |   & {##1} & \mbox{##2} }
  \def\orisopt##1[##2]{
    \\ \left(   & |   & {##1} & \mbox{##2} \right) }
  \begin{array}{rcll}
  #1
  \end{array}
  }
\newcommand{\smallsyntax}[1]{\[\syntaxtable{#1}\]}

\renewcommand\ldots{...}\renewcommand\cdots{{\cdot}{\cdot}{\cdot}}

\newcommand{\initrule}{\mathcal{R}_{init}}
\newcommand{\looprule}{\mathcal{R}_{loop}}
\newcommand{\stoprule}{\mathcal{R}_{stop}}

\newcommand{\tlocalbagname}{\opname{Bag}_l}
\newcommand{\tdistbagname}{\opname{Bag}_d}

\newcommand*{\tlocalbag}[1]{\ensuremath{{\tlocalbagname[{#1}]}}}
\newcommand*{\tdistbag}[1]{\ensuremath{{\tdistbagname[{#1}]}}}
\newcommand*{\tbag}[1]{\opname{Bag}[{#1}]}
\newcommand*{\tset}[1]{\opname{FSet}[{#1}]}
\newcommand*{\tlist}[1]{\opname{List}[{#1}]}
\newcommand\collvar{\textit{Coll}}
\newcommand*{\tcoll}[1]{\collvar[{#1}]}

\newcommand\kwlet{\opname{let}}
\newcommand\kwin{\opname{in}}

\newcommand{\val}{v}
\newcommand{\true}{\opname{True}}
\newcommand{\false}{\opname{False}}
\newcommand{\const}{c}
\newcommand\classname{C}
\newcommand\tuple{\opname{Tuple}}
\newcommand*{\caseclass}[2][\classname]{#1(#2)}

\newcommand{\valtyp}{t} 

\newcommand{\basictyp}{\mathbb{B}} 

\let\funtyp\valtyp 
\newcommand{\functyp}[2]{{#1}\rightarrow{#2}} 

\newcommand*\typeof[1]{\textit{type}(#1)}
\newcommand\sumtyp{+}
\newcommand\subt{\mathrel{<:}}
\newcommand\choicetyp{\mathrel{||}}
\newcommand*{\paramclass}[2][\classname]{#1[#2]}

\newcommand{\pat}{\pi}
\newcommand{\patvar}{a} 
\newcommand*\patmatch[2]{\textrm{m}(#1, #2)}
\newcommand*\matches[3]{\textit{match}(#1, #2)\rightarrow #3}  

\newcommand{\app}[2]{{#1}~{#2}}
\newcommand{\ite}[3]{\opname{if}~{#1}~\opname{then}~{#2}~\opname{else}~{#3}}  
\newcommand{\expr}{e}

\newcommand{\mucriteria}{\textsc{h} }

\newcommand*\matchclosure[1]{\lambda~\langle#1\rangle}
\newcommand*\closure[2]{\matchclosure{#1\to#2}}

\newcommand{\flatmap}[3]{\opname{flmap}(\ensuremath{\lambdaexp{#1}{#2}}, \ensuremath{#3})}

\newcommand{\flatmapf}[2]{\opname{flmap}(\ensuremath{#1}, \ensuremath{#2})}

\newcommand{\cogroup}[2]{\opname{cogroup}(\ensuremath{#1}, \ensuremath{#2})}
\newcommand{\reduce}[3][]{\opname{reduce}(\ensuremath{#2}, \ifstrempty{#1}{\ensuremath{e_{#2}}}{\ensuremath{#1}}, \ensuremath{#3})}
\newcommand{\groupby}[1]{\opname{groupby}(\ensuremath{#1})}

\newcommand{\join}[2]{\opname{join}(\ensuremath{#1}, \ensuremath{#2})}

\newcommand{\reducebykey}[2]{\opname{reduceByKey}(\ensuremath{#1}, \ensuremath{#2})}
\newcommand{\reducebykeyop}{\opname{reduceByKey}}

\newcommand{\fixptn}[2]{\ensuremath{\mu(#1, #2)}}

\newcommand\runmu{R\mu}
\newcommand*{\fixpoint}[3][]{\ensuremath{\mu_{#1} \left({#2}, {#3}\right)}}
\newcommand*{\rfixpoint}[4]{\ensuremath{\runmu_{#1} \left({#2}, {#3}; {#4}\right)}}

\newcommand{\appendToWords}{\opname{appendToWords}}

\newcommand{\env}{\Gamma}
\newcommand{\typinf}[3]{ {#1} \vdash {#2} : {#3}}
\newcommand{\tvartyp}[2]{{#1}[{#2}]}

\newcommand{\evaluatesto}{\leadsto}

\newcommand{\rndfile}[3][rnd]{#1\_#2\_#3}

\newcommand{\rewritesto}{\longrightarrow}

\newcommand{\pushingfilterRule}{\text{PF}}
\newcommand{\semijoinRule}{\text{PJ}}
\newcommand{\pushingaggregRule}{\text{PA}}
\newcommand{\fixpointdistRule}{\text{P}_\text{dist}}
\newcommand{\semijoinfilter}[2][A]{\ifstrempty{#2}{F_A}{F_A(#2)} }
\newcommand{\filtername}[1]{\ifstrempty{#1}{F}{F(#1)}}

\newcommand{\plangloballoop}{\mathcal{P}_1}
\newcommand{\planparallelloop}{\mathcal{P}_2}

\newtheorem{property}{Property}
\theoremstyle{definition}
\newtheorem*{notation}{Notation}

\newenvironment{packeditemize}{%
\begin{itemize}}{\end{itemize}}

\newenvironment{packedenumerate}{%
\begin{enumerate}}{\end{enumerate}}

\newcommand{\mumonoidnoPA}{\mumonoid{}-no-$\pushingaggregRule$}
\newcommand{\mumonoidnoPdist}{\mumonoid-no-$\fixpointdistRule$}

\usepackage{amsthm}

\let\packedsection\section
\let\packedsubsection\subsection
\let\packedsubsubsection\subsubsection

\definecolor{dkgreen}{RGB}{2,112,10}
\definecolor{dkorange}{RGB}{200,120,50}
\begin{document}

\title[Efficient Iterative Programs with Distributed Data Collections]{Efficient Iterative Programs with Distributed Data Collections}

\author[S.Chlyah]{Sarah Chlyah \lmcsorcid{0009-0004-1769-5109}}

\author[N.Gesbert]{Nils Gesbert}

\author[P.Genev\`es]{Pierre Genev\`es \lmcsorcid{0000-0001-7676-2755}}
 
\author[N.Laya\"ida]{Nabil Laya\"ida \lmcsorcid{0000-0001-8472-9365}}

\address{Univ. Grenoble Alpes, CNRS, Inria, Grenoble INP, LIG, 38000 Grenoble}
\email{sarah.chlyah@inria.fr, nils.gesbert@inria.fr, pierre.geneves@inria.fr, nabil.layaida@inria.fr}





\begin{abstract}
  Big data programming frameworks have become increasingly important
  for the development of applications for which performance and
  scalability are critical. In those complex frameworks, optimizing
  code by hand is hard and time-consuming, making automated
  optimization particularly necessary. In order to automate
  optimization, a prerequisite is to find suitable abstractions to
  represent programs; for instance, algebras based on monads or
  monoids to represent distributed data collections. Currently,
  however, such algebras do not represent recursive programs in a way
  which allows for analyzing or rewriting them. In this paper, we extend a
  monoid algebra with a fixpoint operator for representing recursion
  as a first class citizen and show how it enables new optimizations.
  Experiments with the Spark platform illustrate performance gains
  brought by these systematic optimizations.
\end{abstract}


\maketitle{}
\section{Introduction}
With the proliferation of large scale datasets of various data structures (such as graphs, collections, documents, trees, etc.) and in various domains (such as knowledge representation, social networks, transportation, biology, etc.), the need for efficiently extracting information from these datasets becomes increasingly important. This requires the development of methods for effectively distributing both data and computations so as to enable scalability and improve perfomance. Efforts to address these challenges over the past few years have led to various systems such as MapReduce~\cite{dean-osdi04}, Dryad~\cite{dryad2007}, Spark~\cite{ZahariaXWDADMRV16}, Flink~\cite{CarboneKEMHT15}.
While these systems can handle large amounts of data and allow users to write a broad range of applications, 
writing efficient applications is nevertheless not trivial.
Let us consider for instance the problem of finding the shortest paths in a large scale graph. We could write the Spark/Scala program in Fig~\ref{fig:sppgm} to solve it. The \texttt{shortestPaths()} function takes as input a graph \texttt{R} of weighted edges \texttt{(\textcolor{dkorange}{src}, \textcolor{dkorange}{dst}, \textcolor{dkorange}{weight})} and returns the shortest paths between each pair of nodes in the graph.  
The loop (in lines 6 to 14 of Fig~\ref{fig:sppgm}) computes all the paths in the graph and their lengths; to get new paths, edges from the graph get appended to the paths found in the previous iteration using the join operation. Then reduceByKey operation is used to keep the shortest paths. Spark performs the join and distinct operations by transferring the datasets (arguments of the operations) across the workers so as to ensure that records having the same key are in the same partition for join, and that no record is repeated across the cluster for distinct. 
Hence, for optimizing such programs, the programmer needs to take this data exchange into account as well as other factors like the amount of data processed by each worker and its memory capacity, the network overhead incurred by shuffles, etc. One optimization that can be done to reduce data exchange in this program is to assign each worker a part of the graph and make it compute the paths in the graph that start from its own part. This optimization leads to the following program (Fig~\ref{fig:spdistrpgm}.) which is not straightforward to write, less readable, and requires the programmer to give his own local version of dataset operators (such as join) that are going to be used to perform the local computations on each worker.
  \begin{figure*}
  \begin{lstlisting}
  def shortestPaths(R:RDD[(Int,Int,Int)]) = {
    var ret = R
    var X: RDD[(Int, Int, Int)] = R
    var new_cnt = ret.count()
    var cnt = new_cnt
    do {
      cnt = new_cnt
      X = X.map({case (x,y,l1) => (y,(x,l1)) })
        .join(R.map({ case (z,t,l2) => (z,(t,l2)) })) 
        .map({case (_,((x,l1),(t,l2))) => (x,t,l1+l2) })
      ret = ret.union(X).distinct()
      new_cnt = ret.count()
    } while (new_cnt > cnt)
    ret.map({case (x,y,l) => ((x,y),l)}).reduceByKey(min)
  }
\end{lstlisting}
\caption{Shortest paths program.}\label{fig:sppgm}	
\end{figure*}

\begin{figure*}
\begin{lstlisting}
  def shortestPaths(R:RDD[(Int,Int,Int)]) = {
    val dictR = LocalOps.to_dict(((x:(Int,Int,Int)) => x._1),
      (x:(Int,Int,Int)) => x, sc.broadcast(R.collect()).value)
    var r = R.mapPartitions(part => {
      var ret = part.toList
      var X = ret
      var cnt = ret.size
      var new_cnt = cnt
      do {
        count = new_count
        X = LocalOps.join(LocalOps.to_dict(((x:(Int,Int,Int)) => x._2),
            (x:(Int,Int,Int)) => x, X), dictR)
          .map({case (k, ((x,y,l), (a,b,m))) => (x,b,l+m)}) diff ret
        ret = (ret ++ X).distinct
        new_count = ret.size
      } while (new_cnt > cnt)
      ret.toIterator
    })
    r.distinct().map({case (x,y,l) => ((x,y),l)}).reduceByKey(min)
  }
\end{lstlisting}
\caption{Shortest paths program with less data exchange.}\label{fig:spdistrpgm}
\end{figure*}
Another possible optimization is to put the \texttt{reduceByKey} operation inside the loop to keep only the shortest paths at each iteration because each subpath of a shortest path is necessarily a shortest path.
  More generally, finding such program rewritings can be hard. First, it requires guessing which program parts affect performance the most and could potentially be rewritten more efficiently. Second, assessing that the rewriting performs better can hardly be determined without experiments. During such experiments, the programmer might rewrite the program possibly several times, because he has limited clues of which combination of rewritings actually improves performance.

One approach to this problem is to offer the user a Domain Specific Language (DSL) to query the data. A DSL is a high level language that is specialized in a particular application domain, and that can be called from within a general purpose language. Queries in this DSL would be translated to an intermediate representation (e.g. an algebra) so that they can be optimized automatically.
The idea is to relieve users
 from having to worry about optimization in the distributed setting, so that they can focus only on formulating domain-specific queries in a declarative manner.
 A notoriously successful example of this approach is the SQL language and its associated Relational Algebra.
 This success is due to the level of abstraction provided by the declarative syntax of SQL as well as the extensively studied optimizations provided by Relational Algebra. In RA, data is modelled as relations made of rows and columns.
 This means that in order to express complex computations on more complex data like nested collections, using a formalism based on the relational model requires flattening the data and using ad-hoc solutions for supporting user defined functions (UDFs). 
 This means that: 
 (1) At the language level, we could have a query language expressed on a flat data model which causes \emph{impedance mismatch} issues. It is the term that is used to refer to the problems that arise when the data model of the high level language is different from that of the general purpose host language. Specifically, more complex user defined data (data defined by the user in the general purpose programming language) has to be flattened to match the tabular data model of the DSL. In addition, the DSL provides limited support for complex data processing (data transformation, iteration, aggregation, etc.).
 In order to perform custom transformations on data, one could use language extensions like PL/SQL which, in addition to exacerbating the impedance mismatch problem, requires user expertise and provides only limited optimizations. Alternatively, the user could perform data transformations on the query results in the programming language, which increases roundtrips between the program and the database and does not allow for holistic program optimization.
 (2) At the algebraic level, a number of additional joins are introduced to go from hierarchical to flat types and vice versa which has an impact on performance. Additionally, arguments to second order operations are treated as black box functions which means that they cannot be analyzed and transformed to make automatic optimizations.

It is then important to investigate intermediate representations for expressing and optimizing queries that manipulate data in their native format. 
As argued by Meijer in~\cite{Meijer-acm}, establishing and standardizing a formal background for the noSQL market, which now contains multiple separate systems and solutions, is necessary for its economic growth as it was the case for the SQL market thanks to the introduction of RA. The author considers that an algebra based on \emph{monads} is a suitable formalism for this purpose.
Studying intermediate representations that allow for expressing operations on data in their native format would also pave the way for optimizing subsets of general purpose languages and embedded DSLs that do not suffer from impedance mismatch problems.
In the context of big data applications, considered algebras must be able to capture distributed programs on big data platforms and provide the appropriate primitives to allow for their optimization. One example of optimizations is to push computations as close as possible to where data reside.
When programming with big data frameworks, data is usually split into partitions and both data partitions and computations are distributed to several machines. These partitions are processed in parallel and intermediate results coming from different machines are combined, so that a unique final result is obtained, regardless of how data was split initially. This imposes a few constraints on computations that combine intermediate results. Typically, functions used as aggregators must be associative. For this reason, we consider that the monoid algebra is a suitable algebraic foundation for taking this constraint into account at its core.
It provides operations that are monoid homomorphisms, which means that they can be broken down to the application of an associative operator. This associativity implies that parts of the computation can actually be performed in parallel and combined to get the final result.

A significant class of big data programs are iterative or recursive in nature (PageRank, k-means, shortest-path, reachability, etc.). 
Recursion is also a very important feature for graph querying as it enables to navigate through the graph and express traversal queries such as paths of arbitrary length ~\cite{reutter2017, libkinrpq2012, mura-sigmod20}.
Iterations and recursions can be implemented with loops. 
Depending on the nature of the computations performed inside a loop, the loop might be evaluated in a distributed manner or not. Furthermore, certain loops that can be distributed might be evaluated in several ways (global loop on the driver\footnote{In Big Data frameworks such as Spark, the \emph{driver} is the process that creates tasks and sends them to be executed in parallel by \emph{worker} nodes.}, parallel loops on the workers, or a nested combination of the latter). The way loops are evaluated in a distributed setting often has a great impact on the overall program execution cost. Obviously, the task of identifying  which loops of an entire program can be reorganized into more efficient distributed variants is challenging. This often constitutes a major obstacle for automatic program optimization. In the algebraic formalism, having a recursion operator makes it possible to express recursion while abstracting away from how it is executed.

The goal of this work is to introduce a gain in automation of distributed program transformation towards more efficient variants. We focus especially on recursive programs (that compute a fixpoint). For this purpose, we propose an algebra capable of capturing the basic operations of distributed computations that occur in big data frameworks, and that makes it possible to express rewriting rules that rearrange the basic operations so as to optimize the program. 
We build on the monoid algebra introduced in \cite{FegarasN18,fegaras-jfp2017} that we extend with an operator for expressing recursion. This monoid algebra is able to model a subset of a programming language $\mathcal{L}$ (for instance Scala), that expresses computations on distributed platforms (for instance Spark).
\paragraph{Contributions.}
Our contributions are the following:
\begin{packedenumerate}
\item An extension of the monoid algebra with a fixpoint operator. This enables the expression of iteration in a more functional way than an imperative loop and makes it possible to define new rewriting rules;
\item New optimization rules for terms using this fixpoint operator:
\begin{itemize}
\item We show that under reasonable conditions, this fixpoint can be considered as a monoid homomorphism, and can thus be evaluated by parallel loops with one final merge rather than by a global loop requiring network overhead after each iteration; 
\item We also present new rewriting rules with criteria to push filters through a recursive term, for filtering inside a fixpoint before a join, and for pushing aggregations into recursive terms;
\item Finally, we present experimental evidence that these new rules generate significantly more efficient programs.
\end{itemize}
\end{packedenumerate}

\packedsection{The \mumonoid{} Algebra} \label{sec:mumonoidsyntaxsemantics}

In this section, we describe a core calculus, which we call \mumonoid{},
intended to model a subset of a programming language $\mathcal{L}$ (e.g. Scala\footnote{Major Bigdata frameworks like Spark and Flink provide a Scala API and are implemented in Scala which makes Scala a suitable language for our work. Scala also provides reflection which allows generic Scala constructs to be part of the algebra as we will explain later.}) that is used
for computations on a big data framework (through an API provided by the framework). 
\mumonoid{} aims at being as general as possible, while focusing on formalizing computations subject to optimization. It is based upon the monoid algebra of Fegaras~\cite{fegaras-jfp2017}.
Dataset manipulations are
captured as algebraic operations, and specific operations on elements
of those datasets are captured as functional expressions that are
passed as arguments to some of the algebraic operations. In \mumonoid{},
we formalize \emph{some} of those functional constructs, specifically the ones
that we need to analyse in the algebraic expressions. For example, some optimization
rules need to analyse the pattern and body of flatmap expressions in order to check
whether the optimization can take place.

Making explicit only the shapes
that are interesting for the analysis enables to
abstract from the specific programming language $\mathcal{L}$ that we optimize. This way,
constructs of $\mathcal{L}$ other than those which we model
explicitly are represented as \emph{constants} $\const$, as they are
going to be left to $\mathcal{L}$'s compiler to typecheck and
evaluate. We only assume that every constant $\const$ has a type
$\typeof\const$ which is either a basic type or a function type, and
that, when its type is $\functyp{\funtyp_1}{\funtyp_2}$, it can be
applied to any argument of type $\funtyp_1$ to yield results of type
$\funtyp_2$.

We first describe the data model we consider, then in
Sec.~\ref{sec:fegaras} we recall the main definitions of the monoid
algebra proposed by Fegaras \cite{fegaras-jfp2017}. We then introduce
a general notion of \emph{aggregation function} in
Sec.~\ref{sec:aggreg} and our addition to the monoid algebra, the
fixpoint operator $\mu$, in Sec.~\ref{sec:homo}. Then in
Sec.~\ref{sec:syntax} we define the syntax of our own core calculus,
and, in Sec.~\ref{sec:typesys}, a minimal type system for it.
We then proceed to give a denotational semantics for our specific constructs
in Sec.~\ref{sec:densem} and discuss evaluation of expressions in Sec.~\ref{sec:eval}.

\packedsubsection{Data model: distributed collections of data} \label{sec:datamodel}

\subsubsection{Collection monoids}\label{sec:collmon}
We are interested in programs which work on distributed datasets of an
homogenous type. Such a dataset consists in a number of records, which
are all values of the same type, and we assume a cluster of networked machines
where each machine stores some of the records.

Different abstraction levels are possible for such a distributed
dataset. At the programming level, we usually want to abstract away
from the partitioning, i.\,e.\ we consider two states of the storage
as representing the same data if they contain the same records,
regardless of the number of machines and of which machine holds which
records. That way, the program is reasonably independent from the
structure of the cluster it will be run on. We may or may not want to
abstract away from the order in which the records are stored, and we
may or may not want to abstract away from the number of times the same
record appears. Depending of the abstraction level, we thus can see
the dataset as a list, a bag, or a set of records. We regroup finite lists,
finite bags and finite sets under the generic term of
\emph{collections}.

\begin{notation} Given a data type $\valtyp$, and $\collvar$ a
  sort of collection, i.\,e.\ 
  one of $\opname{List}, \opname{Bag}$ or $\opname{FSet}$,
  we write $\tcoll\valtyp$ for the set of collections of the sort
  $\collvar$ containing values of type $\valtyp$.
\end{notation}

Let us now recall the algebraic definition of a \emph{monoid}:
\begin{defi}
  A \emph{monoid} is a triple $(S, \otimes, e)$ where $S$ is a set,
  $\otimes$ an associative binary operation on $S$, and $e$ a neutral
  element for $\otimes$, i.\ e.\ such that:
  \begin{align*}
    \forall x,y,z\in S\quad x\otimes(y \otimes z) &= (x\otimes y) \otimes
                                               z\\
    \forall x\in S\quad x\otimes e  = &x = e\otimes x
  \end{align*}
\end{defi}

As noticed by Fegaras \cite{fegaras-jfp2017}, our three sorts of
collections are particularly useful for representing distributed data
because they each have the algebraic structure of a monoid, where the
neutral element is the empty collection and the associative operator
is respectively list union (i.\,e.\ concatenation) $\lunion$, bag
union $\bunion$ and set union $\sunion$.
Associativity means that the whole collection can be
seen as the union of the subcollections stored on the different
machines without specifying an order in which to apply the union
operator.

The three sorts of \emph{collection monoids}, as Fegaras terms them,
can be related with equivalence relations, reflecting the fact that
they represent different abstraction levels for the same data. To
formalize this, we recall the algebraic definitions of congruence and
quotient monoid:
\begin{defi}[Congruence]
  Let $(A, \otimes, e)$ be a monoid. Let $\sim$ be an equivalence
  relation on $A$. We say that $\sim$ is a \emph{congruence} on the
  monoid if it is
  \emph{compatible} with $\otimes$, i.\ e.\ if:
  $$\forall a, b, a', b'\quad a\sim a'\wedge b\sim b' \implies a\otimes
  b\sim a'\otimes b'$$
\end{defi}
\begin{defi}[Quotient monoid]
  Let $(A, \otimes, e)$ be a monoid and $\sim$ a congruence on it. For
  $a\in A$, let $\hat a = \{b\in A\mid b\sim a\}$, the equivalence class of $a$. Let
  $\hat A = \{\hat a\mid a\in A\}$ be the set of all equivalence
  classes, and let
  $\hat\otimes$ be the operation on $\hat A$ defined by $\hat a\mathbin{\hat\otimes}\hat
  b\eqdef \widehat{a\otimes b}$. This is well-defined because $\sim$ is
  compatible with $\otimes$, so that the result is the same
  independently of the particular choice of $a$ and $b$ in their
  equivalence class.

  Then $(\hat A, \hat\otimes, \hat e)$ is a monoid, termed the
  \emph{quotient monoid} of $(A, \otimes, e)$ by $\sim$ and noted $(A,
  \otimes, e)/\sim$.
\end{defi}

Let
$\eqcomm$ be the congruence on $(\tlist\valtyp, \lunion,
\emptylist)$ generated
by commutativity, i.\,e.\ the smallest congruence such that:
$\forall a, b\ a\lunion b \eqcomm b\lunion a$.
This relation relates all lists containing exactly the same elements,
with the same multiplicity, in any order. So a \emph{bag} can be seen
as an equivalence class of lists for $\eqcomm$, meaning that
$(\tbag\valtyp, \bunion, \emptybag)$ is the quotient monoid
$(\tlist\valtyp, \lunion, \emptylist)/\eqcomm$.

Similarly, let
$\eqidem$ be the congruence on bags generated by idempotence
($a\bunion a\eqidem a$): it relates all bags containing the same
elements, regardless of their multiplicity, and we have that
$(\tset\valtyp, \sunion, \emptyset)$ is the quotient monoid
$(\tbag\valtyp, \bunion, \emptybag)/\eqidem$.

In this work, we choose bags as the default base abstraction level,
since there is no canonical ordering of the machines in the cluster; but
this can be adapted to work with lists. So from now on, we consider
that a dataset is a distributed bag. We now define the formal syntax
of our data model before developing further how we can sometimes work
up to equivalence relations if, e.\ g., we are in fact interested in
sets and not bags.

\subsubsection{Values and types of  \mumonoid{}}
In order to enable algebraic datatypes, we assume an infinite set of \emph{constructors} $\classname$ which can be applied to any number of values. We assume this set contains the special constructors \true, \false{} and \tuple{} for which we will define some syntactic sugar.

The syntax of considered data values is defined as follows:
\begin{small}
\smallsyntax{      
\singleentry \val [] 
	  \const      [constant]
          \oris \caseclass{\val_1, \val_2, \ldots, \val_n} [$n$-ary constructor]
      \oris \bag{\val_1,\ldots,\val_n} [bag]
}
\end{small}%
As mentioned previously (Sec.~\ref{sec:mumonoidsyntaxsemantics}), a constant $c$ can be any value from the language $\mathcal{L}$ (in particular any function) that is not explicitly defined in our syntax.

We define the following syntax for types:  
\begin{small}
  \[
    \begin{array}{rcllrcll}
\valtyp_l & \syntaxdef & & \!\!\!\! \text{local type}

& \valtyp & \syntaxdef & & \text{type}\\
          && \basictyp & \text{~~basic type} && | & \valtyp_l \\
      & | &  {\paramclass[\classname_1]{\valtyp_l,\ldots,\valtyp_l} \choicetyp\cdots\choicetyp \paramclass[\classname_n]{\valtyp_l,\ldots,\valtyp_l}} & \text{~~sum type} && | & \tdistbag{\valtyp_l} & \text{~~~~distributed\ bag type}\\ 
          & | & \tlocalbag{\valtyp_l} & \text{~~local\ bag type~~~~} && | & \functyp{\valtyp}{\valtyp} & \text{~~~~function\ type}\\
    \end{array}
  \]
\end{small}

where $\basictyp$ represents any arbitrary basic type (i.e., considered as a constant atomic type in our formalism). 

In sum types, all constructors have to be different and their order is
irrelevant. They represent values which can belong to any of the case
types $\paramclass[\classname_1]{\valtyp_l,\ldots,\valtyp_l}
\ldots \paramclass[\classname_n]{\valtyp_l,\ldots,\valtyp_l}$ and can
be deconstructed by pattern-matching.

We also define product types $\valtyp_1 \times\cdots\times \valtyp_n$
as syntactic sugar for
$\paramclass[\tuple]{\valtyp_1,\ldots,\valtyp_n}$, i.\,e.\ a
particular case of constructor type.

For a given type $t$, we denote by $\tlocalbag{t}$ the type of a local
bag and by $\tdistbag{t}$ the type of a distributed bag of values of
type $t$. Notice that we can have distributed bags of any data type
$t$ including local bags, which allows us to have nested collections.
We allow data distribution only at the top level though (distributed
bags cannot be nested).

An important feature of Fegaras’ monoid algebra, and of \mumonoid{},
is that all algebraic operations are defined in a way which is
agnostic to distribution. So, although we introduce the distinction
between $\tlocalbag\valtyp$ and $\tdistbag\valtyp$ in order to prevent
nesting distributed bags, we will use the notation $\tbag{t}$ to
represent a bag which may or may not be distributed when both are
possible and it does not affect the semantics.

\subsection{Fegaras’ monoid algebra}\label{sec:fegaras}
In this section, we recall briefly the main definitions from Fegaras’
monoid algebra \cite{fegaras-jfp2017}, upon which our work is based.

The monoid algebra is based on the three sorts of collection monoids
described in \ref{sec:collmon} and on \emph{collection homomorphisms}.

We first recall the definition of a monoid homomorphism:
\begin{defi}
  Let $(A, \otimes, e)$ and $(B, \odot, \varepsilon)$ be two monoids. A
  \emph{monoid homomorphism} from $A$ to $B$ is a function $h$ from $A$ to
  $B$ such that:
  \begin{align*}
    h(e) &= \varepsilon \text{ and}\\
    \forall x, y\in A \quad h(x\otimes y) &= h(x)\odot(y).
  \end{align*}
\end{defi}

\emph{Collection homomorphisms} are now
defined using the following universal property enjoyed by the
collection monoids (which are \emph{free structures} in the algebraic sense):
\begin{property}[universal property of collection monoids]\label{prop:universalproperty}
For $\collvar$ a collection monoid, let $\mathbb{U}_\collvar$ be the
corresponding singleton construction function. Let $(A, \otimes, e)$
be a monoid which satisfies all the algebraic laws of $\collvar$
(i.\ e.\ commutativity for bags, and commutativity and idempotence
for sets). Let $f : \valtyp \to A$ be a function.

Then there exists a unique monoid homomorphism
$\homo{\otimes}: \tcoll\valtyp\to A$ such that:
$\homo{\otimes}(\mathbb{U}_\collvar(x)) = f(x)$.
\end{property}
This \emph{collection homomorphism}
applies the function $f$ to all the elements of the input collection
and combines all the results together with the operation $\otimes$,
yielding a single element of $A$.
 
Fegaras’ monoid algebra comprises a number of collection
homomorphisms, all defined in the form $\homo{\otimes}$ for
appropriate $f$s and $\otimes$s. We refer the reader to
\cite{fegaras-jfp2017} for the detail.
In the present work, we use a slightly different set.
Namely:
\begin{itemize}
\item we do not consider orderBy because we concentrate on bags
rather than lists;
\item we use reduceByKey rather than groupBy (together with the other
  operations, they lead to the same expressivity);
\item we add the join operation. Even though it can be expressed in
  terms of coGroup, this operation is useful for us to have as a
  primitive because it is an
  homomorphism from each of its two arguments separately, whereas coGroup is only a
  binary homomorphism (an homomorphism from the product monoid of its
  two arguments).
\end{itemize}

Our set of primitive operations is thus the following,
here presented in a way adapted to our default abstraction level of bags:

\begin{itemize}
\item $\flatmapf{f}{X}$, with $f : \valtyp_1\to\tlocalbag{\valtyp_2}$ and $X :
  \tbag{\valtyp_1}$ is the flatmap operation: it applies $f$ to each
  element of $X$ and merges all the results into a single bag using
  bag union $\bunion$. This operation is a monoid homomorphism from
  $\tbag{\valtyp_1}$ to $\tbag{\valtyp_2}$, so that if $X$ is distributed it
  can be run separately on each local subcollection without any data
  exchange. Note the restriction that $f$ is not allowed to return
  distributed bags. It makes the flatmap operator less general than the
  mathematical function it represents but reflects what we have in
  distributed data frameworks.
\item $\reduce{\oplus}{X}$, with $X : \tbag\valtyp$ and $(\valtyp,
  \oplus, e_\oplus)$ a commutative monoid\footnote{Commutativity is needed 
  because the monoid of bags is commutative (see Proposition~\ref{prop:universalproperty})}, 
  reduces the input dataset by combining all its elements with $\oplus$. 
  For example: ${\reduce[0]{+}{\bag{1,4,6}}} = 11$. This operation is a monoid
  homomorphism from $\tbag\valtyp$ to $(\valtyp, \oplus, e_\oplus)$,
  so that if $X$ is distributed it can be run separately on each local
  subcollection before combining all the local results once.
\item $\reducebykey{\oplus}{X}$, with $X :
  \tbag{\valtyp_1\times\valtyp_2}$ and $\oplus$ an associative and
  commutative binary operation on $\valtyp_2$,
  takes as argument a bag of elements in the form $(k, v)$ (key-value
  pair) and combines all values $v$ having the same key $k$ into a
  single one using the $\oplus$ operator. For example:
  $\reducebykey{+}{\bag{(1,2), (1, 4), (2,2), (2,1), (1,3)}} =
  \bag{(1,9), (2,3)}\}$. This operation is a monoid
  homomorphism which can again be run
  separately on each local subcollection before combining the results.
\item $\join{X}{Y}$, with $X : \tbag{\valtyp_1\times\valtyp_2}$ and $Y
  : \tbag{\valtyp_1\times\valtyp_3}$, is
  the join-by-key operation: it takes two collections of elements of
  the form $(k, v)$ and $(k, w)$, and returns a collection of elements
  of the form $(k, (v, w))$, one for each pair $(v, w)$ of values
  having the same key $k$. If a key appears $n$ times in one input
  dataset and $m$ times in the other, it appears $nm$ times in the
  result. It is a monoid homomorphism from each of its arguments to
  $\tbag{\valtyp_1\times(\valtyp_2\times\valtyp_3)}$, so that if any
  of the input bags is distributed it can be run separately on each
  local subcollection for that one.

  Note that, algebraically, the join operation can be written with
  flatmaps (this is a feature of all homomorphisms from bags to bags);
  however, if both inputs are distributed then this is not possible in
  \mumonoid{} without violating the restriction on the functional
  argument of flatmap, which justifies including join as a primitive.

\item $\cogroup{X}{Y}$, with $X : \tbag{\valtyp_1\times\valtyp_2}$ and $Y
  : \tbag{\valtyp_1\times\valtyp_3}$, takes two collections of elements of the form
  $(k, v)$ and $(k,w)$ and returns a collection of elements of the
  form $(k, (V, W))$ where $V$ and $W$ are the sets of v values and w
  values having the same key $k$. This operation is an homomorphism
  from the product monoid
  $\tbag{\valtyp_1\times\valtyp_2}\times\tbag{\valtyp_1\times\valtyp_3}$
  to $\tbag{\valtyp_1\times(\tlocalbag{\valtyp_2}\times\tlocalbag{\valtyp_3})}$.
\end{itemize}

Additionnally, Fegaras’ monoid algebra includes a \textsf{repeat}
operation, which is not an homomorphism. In this work, we replace this
operation with our own $\mu$ operation, explained in detail in
Section~\ref{sec:homo}. Before we get there, we introduce
our notion of \emph{aggregation function}.

\subsection{Equivalence relations and aggregation functions}\label{sec:aggreg}
It is quite common in practice that a programmer is only
interested in the set of values of a dataset, not in potential
duplicates the bag representing the storage may contain. So this
programmer will work with bags up to $\eqidem$ (see
Sec.~\ref{sec:collmon}).
We can notice that each equivalence class of bags has a canonical representant: the bag
where each element appears only once. Let $\distinct:
\tbag\valtyp\to\tbag\valtyp$ be the function which removes duplicates,
returning this canonical representant. This function is useful, but
costly to compute in a distributed context, since duplicates can occur
across different machines and eliminating them thus involves a lot of
communication over the network. Therefore, it should not be used all
the time but only when necessary: bags with duplicates can be used in
intermediary computation steps, where we tolerate redundant
information temporarily.

Sometimes, the programmer is not even interested in the whole set of
values, but only in more synthetic information about the dataset. For
example, in the shortest path problem: if we have a dataset containing
paths together with their length and this dataset contains two paths
from $a$ to $b$ with different lengths, then the longer path is
irrelevant and can be considered redundant even though it is not the
same value as the other one. It is useful to also think of this
situation in terms of an equivalence relation: two bags are equivalent
for this purpose iff they contain exactly the same \emph{shortest}
paths. Then the canonical representant of an equivalence class is the
bag with no duplicates which does not contain any non-shortest path.
We can also see that the function $\delta$ which removes non-shortest
paths (and duplicates) from a dataset has features analogous to
$\distinct$, as we will detail below. We regroup such functions under
the term \emph{aggregation functions}.

\begin{defi}[aggregation function]\label{def:aggreg}
  We call \emph{aggregation function} any function $\delta:
  \tbag\valtyp\to\tbag\valtyp$ with the following properties:
  \begin{itemize}
  \item $\delta(\emptybag) = \emptybag$
  \item $\forall a, b\quad\delta(a\bunion b) = \delta(\delta(a)
\bunion \delta(b))$
  \end{itemize}
Note that these two properties also imply that $\delta$ is idempotent.
\end{defi}
Remark that our definition excludes some functions which could be
considered aggregators in a more general sense, e.\ g.\ functions
computing an average.
\begin{defi}
The equivalence relation associated with an aggregation function,
$\eqdelta$, is defined by:
$$a \eqdelta b \equivdef \delta(a) = \delta(b)$$
\end{defi}
\begin{lem}
  Let $\delta$ be an aggregation function, then $\eqdelta$ is
  compatible with the monoid operation $\bunion$, i.\,e.\ we have:
$$\forall a, b, a', b'\quad a\eqdelta a'\wedge b\eqdelta b' \implies
a\bunion b\eqdelta a'\bunion b'$$
\end{lem}
\begin{proof}
  We have $\delta(a\bunion b) = \delta(\delta(a)\bunion\delta(b)) =
  \delta(\delta(a')\bunion\delta(b')) = \delta(a'\bunion b')$.
\end{proof}
\begin{defi}
  Let $\delta$ be an aggregation function, we define the binary
  operation $\opdelta$ on bags as follows:
$$a \opdelta b \eqdef \delta(a\bunion b)$$
\end{defi}
\begin{lem}\label{lem:aggreghomo}
  Let $\delta(\tbag\valtyp)$ be the image of $\delta$.
  Then $(\delta(\tbag\valtyp), \opdelta, \emptybag)$ is a monoid, noted
  $\mondelta$, isomorphic to the quotient monoid
  $\tbag\valtyp/\eqdelta$, and $\delta$ is a monoid homomorphism:
  $\tbag\valtyp\to\mondelta$.
\end{lem}
\begin{proof}
  Since $\delta$ is idempotent, we can consider $\delta(a)$ the
  canonical representant of the equivalence class of $a$; then we have
  an isomorphism between the equivalence classes (the monoid
  $\tbag\valtyp/\eqdelta$) and their canonical representants (the
  monoid $\mondelta$).
\end{proof}
\begin{exa}
  The function $\distinct: \tbag\valtyp\to\tbag\valtyp$
  which removes duplicates from a bag is an
  aggregation function; $\eqdelta[\distinct]$ is the relation
  $\eqidem$; $\opdelta[\distinct]$ is distinct union of bags
  $\sunion$; and $\mondelta[\distinct]$ is the monoid of bags with no
  duplicates, isomorphic to $\tset\valtyp$.
\end{exa}

Finally, in order to work up to equivalence relations, we need the
notion of compatibility between an homomorphism $\phi$ from bags to
bags and an aggregation function $\delta$:
\begin{lem}
    Let $\phi:\tbag\valtyp\to\tbag\valtyp$ be a monoid homomorphism and
    $\delta:\tbag\valtyp\to\tbag\valtyp$ an aggregation function.
  The following three properties are equivalent:
  \begin{enumerate}
    \item $\forall a, b\quad a\eqdelta b \implies \phi(a)\eqdelta\phi(b)$
    \item $\forall a, b\quad \phi(a\opdelta b) \eqdelta \phi(a)\opdelta\phi(b)$
    \item $\delta\circ\phi\circ\delta = \delta\circ\phi$      
  \end{enumerate}
\end{lem} 
\begin{proof}
  Assume (1) is true. Let $a$ and $b$ be any bags. We have
  $\phi(a)\opdelta\phi(b) = \delta(\phi(a)\bunion\phi(b)) =
  \delta(\phi(a\bunion b))$ (because $\phi$ is a homomorphism). We
  also have $a\bunion b \eqdelta \delta(a\bunion b)$, by definition of
  $\eqdelta$ since $\delta$ is idempotent.
  Therefore, using (1), $\phi(a\bunion b)\eqdelta \phi(\delta(a\bunion
  b))$, and this last term is $\phi(a\opdelta b)$; hence (2).

  Assume (2) is true. Let $a$ be any bag, by taking for $b$ the empty
  bag and using the definitions, (2) yields
  $\delta(\phi(\delta(a\bunion\emptybag))) =
  \delta(\phi(a)\bunion\phi(\emptybag))$. Since $\phi$ is an
  homomorphism we have $\phi(\emptybag) = \emptybag$, thus
  $\delta(\phi(\delta(a))) = \delta(\phi(a))$; hence (3).

  Assume (3) is true. Let $a$ and $b$ such that $a\eqdelta b$. Using
  (3) and the definition of $\eqdelta$, we have
  $\delta(\phi(a)) = \delta(\phi(\delta(a))) = \delta(\phi(\delta(b)))
  = \delta(\phi(b))$; hence (1).
\end{proof}
\begin{defi}\label{def:compat}
  We say that $\phi$ is \emph{compatible} with $\delta$ if any of
  these properties is true. Note that (2) can also be formulated as:
  $\delta\circ\phi$ is a monoid homomorphism from $\mondelta$ to $\mondelta$.
\end{defi}

This definition is strongly related to the premappability condition in
Datalog \cite{zaniolo-tplp17}, as made more apparent by property (3).

\begin{property}
  $\distinct$ is compatible with all homomorphisms $\phi:\tbag\valtyp\to\tbag\valtyp$.
\end{property}
\begin{proof}
  In the following, we write $m\cdot X$, where $X$ is a bag, to denote $X$ combined $m$
  times with itself using $\bunion$.

  Let $A$ be a finite bag of values of type $\valtyp$. Let
  $a_1,\ldots,a_n$ be the distinct values it contains and
  $m_1,\ldots,m_n$ the number of times each one appears in the bag.
  Since $\phi$ is an homomorphism, we have $\phi(A) = \biguplus_{1\leq
  i\leq n}m_i\cdot\phi(\singleton{a_i})$. Then, $\distinct(\phi(A)) =
\bigcup_{1\leq i\leq n}\distinct(\phi(\singleton{a_i}))$: this is
independent from the $m_i$ (since $\sunion$ is idempotent). Thus,
$\distinct(\phi(A)) = \distinct(\phi(\distinct(A)))$.
\end{proof}

\subsection{The $\mu$ operation}\label{sec:homo}
Our purpose is to extend the monoid algebra (as defined in
\ref{sec:fegaras}) with an operator for expressing
iteration which allows effective optimisations when working with a
distributed dataset (note that this does not preclude more general
loops outside our algebra).

As a first idea, consider the following type of iteration.
Let $\phi : \tbag{\valtyp}\to\tbag{\valtyp}$ be a
monoid homomorphism. We start with a bag $R$, then:
\begin{enumerate}
\item at each iteration, $\phi$ is executed on the result
  of the previous iteration;
\item the results of all iterations are accumulated into a single bag;
\item it ends when $\phi$ adds nothing to the results; then the
  bag of all results is returned.
\end{enumerate}
Algebraically, this amounts to computing $\biguplus_{n\in\N}\phi^n(R)$.
The fact that $\phi$ is a monoid homomorphism implies that, if $R$ is
distributed, such a loop can be executed separately on each sub-bag,
with no communication necessary, which is very good for efficiency.
However, if in fact we are not interested in bags with duplicates but
only in sets, i.\,e.\ if we work up to $\eqidem$, it has the serious
drawback that it only stops when $\phi$ returns the empty bag: this
can prevent termination in cases where $\phi$ generates nothing really
new but adds indefinitely more duplicates. A typical example is if we
want to compute the transitive closure of a relation with cycles.

Therefore, it makes sense to periodically remove duplicates. However,
it may not be necessary to remove them \emph{globally} (which is
costly as it involves network communication), as we will detail in
Section~\ref{sec:distfixpoint}. More generally, we can
add to the loop, as a parameter, an aggregation function $\delta$ to
be run at each iteration step. Our general iteration operator $\mu$ is thus:

\begin{defi}[$\mu$ operator]\label{def:mu}
  Let $\phi : \tbag{\valtyp}\to\tbag{\valtyp}$ be a monoid homomorphism,
  $\delta : \tbag{\valtyp}\to\tbag{\valtyp}$ an aggregation function,
  and $R : \tbag{\valtyp}$ a dataset. Assume $\phi$ and $\delta$ are
  compatible (Def.~\ref{def:compat}). The operation
  $\fixpoint[\delta]{R}{\phi}$ computes the following sequences:
  \begin{itemize}
  \item $R_0 = \delta(R)$
  \item $S_0 = R_0$
  \item $R_{n+1} = \delta(\phi(R_n))$
  \item $S_{n+1} = S_n\opdelta R_{n+1}$
  \end{itemize}
  until it reaches an $N$ such that $S_{N+1} = S_N$; then it returns
  $S_N$.
\end{defi}

Note that the first idea discussed before is the particular case where
$\delta$ is the identity function. Also note that the requirement that
$\delta$ and $\phi$ are compatible is automatically true when $\delta$
is either the identity function or $\distinct$.

In the following, we consider $\distinct$ the default aggregation
function and write $\fixpoint{R}{\phi}$ as a shortcut for
$\fixpoint[\distinct]{R}{\phi}$. Our idea is that programmers would
usually not write $\mu$ terms with a different $\delta$ themselves,
but they can be obtained through rewriting and used for optimization
(Sec.~\ref{sec:pushagg}).

\subsection{Syntax of $\mu$-monoids}\label{sec:syntax}

The syntax of expressions is formally defined as follows:
\begin{small}
  \smallsyntax{
    \singleentry \pat []
    \patvar \mid \caseclass{\pat_1, \pat_2, \ldots, \pat_n} [pattern: variable, constructor pattern]\\
    \singleentry \expr []
\const \mid \patvar \mid \singleton{\expr} [expression: constant, variable, singleton]
\oris \matchclosure{\pat_1\to \expr_1\mid\cdots\mid\pat_n\to\expr_n} [function with pattern matching]
\oris \app{\expr}{\expr}  \mid \caseclass{\expr_1,\expr_2,\ldots,\expr_n} [application, constructor expression]
\oris \flatmapf{\expr}{\expr}
\mid {\reduce[\expr]{\expr}{\expr}}
[flatmap, reduce]
\oris \reducebykey{\expr}{\expr}
\mid \cogroup{\expr}{\expr}
\mid \join{\expr}{\expr}\quad [reduce by key, cogroup, join by key]
\oris {\fixpoint[\expr]{\expr}{\expr}} [fixpoint]
}
\end{small}

To this, we add the following as syntactic sugar:
\begin{packeditemize}
\item $(e_1,\ldots,e_n)$ with no constructor is an abbreviation for: $\caseclass[\tuple]{e_1, \ldots, e_n}$
\item $\ite{\expr}{\expr_1}{\expr_2}$ is an abbreviation for:
  $\app{\matchclosure{\true\to\expr_1\mid\false\to\expr_2}}{\expr}$,
  i.\,e.\ a particular case of pattern-matching against the two
  constant constructors $\true$ and $\false$ representing Boolean values.
\item $\groupby\expr$ is an abbreviation for:
  $\reducebykey{\bunion}{\flatmap{(k, v)}{(k, \singleton{v})}{\expr}}$
\item Constants $\const$ can also represent functions (defined in the 
language $\mathcal{L}$). We consider operators such as the bag union operator
$\bunion$ as constant functions of two arguments and use the infix
notation as syntactic sugar.
\item To make examples more readable, we use the $\kwlet~\textbf{name}
  = \expr_1\ \kwin\ \expr_2$ syntax with the usual meaning.
\end{packeditemize}

\paragraph{Example:}
\begin{small}\begin{align*}
  &\kwlet~\text{appendToWords} = \closure{X}{{\flatmap{x}{
  \flatmap{c}{\ite{\opname{contains}\,x\,c}{\singleton{}}{\singleton{x+c}}}{C}
}{X}}}\\&\kwin\ \fixptn{C}{\text{appendToWords}}\end{align*}
\end{small}%

This expression computes the set of all possible words with no
repeated letters that can be formed from a set of characters C. We
assume that $+$ and $\opname{contains}$ are defined in $\mathcal{L}$:
$+$ appends its second argument to the first and $\opname{contains}$ checks whether the first argument is contained in the second argument.
$\appendToWords$ thus returns a new set of words from a given set of
words $X$ by appending to each of the words in $X$ each letter in $C$
whenever it was not already present.

The iteration operator computes the following, where we consider $C =
\bag{a,b,c}$ --- the fixpoint is reached in 3 steps:
\begin{small}
\begin{flalign*}  
&R_0 = C && S_0 = C\\
&R_1 = \distinct(\phi(C)) = \bag{ab,ac,ba,bc,ca,cb}
&&S_1 = C\cup R_1 = \bag{ab,ac,ba,bc,ca,cb, a, b, c}\\
&R_2 = \bag{abc,acb,bac,bca,cab,cba}
&&S_2 = \bag{abc,acb,bac,bca,cab,cba,ab,ac,ba,bc,ca,a,b,c}\\
&R_3 = \distinct(\phi(R_2)) = \bag{} && S_3 = S_2\cup R_3 = S_2
\end{flalign*}%
\end{small}%

\subsection{Well-typed terms}\label{sec:typesys}

We define typing rules for algebraic terms, in order to exclude
meaningless terms.
In these rules, we use \emph{type environments} $\Gamma$ which bind variables to
types. An environment contains at most one binding for a given
variable. We combine them in two different ways:
\begin{packeditemize}
\item $\Gamma\cup\Gamma'$ is only defined if $\Gamma$ and $\Gamma'$
  have no variable in common, and is the union of all bindings in
  $\Gamma$ and $\Gamma'$;
\item $\Gamma + \Gamma'$ is defined by taking all bindings in
  $\Gamma'$ plus all bindings in $\Gamma$ for variables not appearing in
  $\Gamma'$. In other words, if a variable appears in both, the
  binding in $\Gamma'$ overrides the one in $\Gamma$.
\end{packeditemize}

\begin{defi}[matching]
We first define the environment obtained by \emph{matching} a data type to a
pattern by the following: %
\begin{small}
$$\inferrule{~}{\matches{\patvar}{\funtyp}{\patvar : \funtyp}}$$
$$\inferrule{\forall i\ \matches{\pat_i}{\valtyp_i}{\Gamma_i}}
  {\matches{\caseclass{\pat_1,\ldots,\pat_n}}{\paramclass{\valtyp_1,\ldots, \valtyp_n}}{\Gamma_1\cup\cdots\cup\Gamma_n}}$$
\end{small}
If, according to these rules, there is no $\Gamma$ such that
$\matches{\pat}{\funtyp}{\Gamma}$ holds, we say that pattern $\pat$ is
\emph{incompatible} with type $\funtyp$. Note that, with our conditions, a pattern
containing several occurrences of the same variable is not compatible
with any type and hence cannot appear in a well-typed term, as the
typing rules will show.
\end{defi}

In order to type functions with pattern-matching, we define the following
operation for combining sum types:

\begin{defi}
The operation $\sumtyp$ on sum types is defined recursively as follows. Let $t$ be a sum type and $\classname$ a constructor not appearing in $t$, then:
\begin{small}
\begin{align*}
  t \sumtyp (t'_1\choicetyp\cdots\choicetyp t'_m) &= (t \sumtyp t'_1) \sumtyp (t'_2\choicetyp\cdots\choicetyp t'_m)\\
 t \sumtyp \paramclass{\valtyp_1,\ldots,\valtyp_n} &= t \choicetyp \paramclass{\valtyp_1,\ldots,\valtyp_n}\\
(t\choicetyp\paramclass{\valtyp_1,\ldots,\valtyp_n}) \sumtyp \paramclass{\valtyp_1',\ldots,\valtyp_n'} &= t\choicetyp\paramclass{\valtyp_1+\valtyp_1',\ldots,\valtyp_n+\valtyp_n'}
\end{align*}%
\end{small}%
If $t$ is not a sum type, we define $t\sumtyp t = t$.
The type $t\sumtyp t'$ is not defined if $t\neq t'$ and $t$ or $t'$ is not a sum type, or if they have constructors in common with incompatible type parameters, i.\,e.\ type parameters which cannot themselves be combined with $\sumtyp$.
\end{defi}

\begin{defi}[Subtyping]
We define subtyping as follows (it is nontrivial only for sum types):
$$t\subt t' \equivdef t\sumtyp t' = t'$$
\end{defi}

\begin{defi}[Well-typed terms]
A term $\expr$ is
\emph{well-typed} in a given environment $\Gamma$ iff
$\typinf{\env}{\expr}{\valtyp}$ for some type
$\valtyp$, as judged by the relation defined in
Figure~\ref{fig:typesystem}. In these rules, $T$ represents one of
$\tlocalbagname$ or $\tdistbagname$.
\end{defi}

Note that these rules do not give a way
to infer the parameter type of a $\lambda$ expression in general; we
assume some mechanism for that in the language $\mathcal{L}$.

\begin{figure*}
\begin{small}
\begin{mathpar}
\inferrule{ \typinf{\env}{\expr_1}{\functyp{\valtyp}{\tlocalbag{\valtyp'}}}
  \and \typinf{\env}{\expr_2}{\tvartyp{T}{\valtyp}}} 
{ \typinf{\env}{\flatmapf{\expr_1}{\expr_2}}{\tvartyp{T}{\valtyp'}} }

\inferrule{
  \typinf{\env}{\expr_1}{\functyp{\functyp\valtyp\valtyp}{\valtyp}}
  \and \typinf{\env}{\expr_2}{\valtyp}
       \and \typinf{\env}{\expr_3}{\tvartyp{T}{\valtyp}}
     }
{ \typinf{\env}{\reduce[\expr_2]{\expr_1}{\expr_3}}{t} }



\inferrule{ \typinf{\env}{\expr_1}{\functyp{\functyp{\valtyp'}{\valtyp'}}{\valtyp'}}
  \and \typinf{\env}{\expr_2}{\tvartyp{T}{\valtyp\times\valtyp'}}}
{\typinf{\env}{\reducebykey{\expr_1}{\expr_2}}{\tvartyp{T}{\valtyp\times\valtyp'}}}

\inferrule{  \typinf{\env}{\expr_1}{\tvartyp{T_1}{\valtyp \times \valtyp_1}} \and
			 \typinf{\env}{\expr_2}{\tvartyp{T_2}{\valtyp \times \valtyp_2}} \and 
             T_3 = (\ite{T_1=T_2}{T_1}{\tdistbagname})
             }
{ \typinf{\env}{\cogroup{\expr_1}{\expr_2}}{\tvartyp{T_3}{\valtyp \times \left( \tlocalbag{\valtyp_1} \times \tlocalbag{\valtyp_2} \right)}} }

\inferrule{  \typinf{\env}{\expr_1}{\tvartyp{T_1}{\valtyp \times \valtyp_1}} \and
			 \typinf{\env}{\expr_2}{\tvartyp{T_2}{\valtyp \times \valtyp_2}} \and 
             T_3 = (\ite{T_1=T_2}{T_1}{\tdistbagname})
             }
{ \typinf{\env}{\join{\expr_1}{\expr_2}}{\tvartyp{T_3}{\valtyp \times \left( \valtyp_1 \times \valtyp_2 \right)}} }

\inferrule{ \typinf{\env}{\expr_1}{\tvartyp{T}{\valtyp}}
        \and
        \typinf{\env}{\expr_2}{\functyp{\tvartyp{T}{\valtyp}}{\tvartyp{T}{\valtyp}}}
        \and \typinf{\env}{\expr}{\functyp{\tvartyp{T}{\valtyp}}{\tvartyp{T}{\valtyp}}}
      }
{ \typinf{\env}{\fixpoint[\expr]{\expr_1}{\expr_2}}{\tvartyp{T}{\valtyp}} }

\inferrule{\forall i\ \typinf{\env}{\expr_i}{t_i} }
{ \typinf{\env}{\caseclass{\expr_1,\expr_2,\ldots, \expr_n}}{\paramclass{t_1,t_2,\ldots, t_n}} }

\inferrule{\typinf{\env}{\expr}{t} }
{ \typinf{\env}{\singleton{\expr} }{\tlocalbag{t}} }

\inferrule{ \valtyp'_1\sumtyp\cdots\sumtyp\valtyp'_n = t' \and \matches{\pat_i}{\valtyp'_i}{\env'_i} \and \typinf{\env+\env'_i}{\expr_i}{\valtyp_i} \and \valtyp_1\sumtyp\cdots\sumtyp\valtyp_n = \valtyp}
{ \typinf{\env}{\matchclosure{\pat_1\to\expr_1\mid\cdots\mid\pat_n\to\expr_n}}{\functyp{\valtyp'}{\valtyp}} }

\inferrule{ \typinf{\env}{\expr_1}{\functyp{\valtyp_1}{\valtyp'}}
        \and \typinf{\env}{\expr_2}{\valtyp_2} \and \valtyp_2\subt\valtyp_1}
{ \typinf{\env}{\app{\expr_1}{\expr_2}}{\valtyp'} }

\inferrule{ \env(\patvar) = \funtyp}
{ \typinf{\env}{\patvar}{\funtyp} }

\inferrule{ \typeof\const = \funtyp }
{\typinf\env\const\funtyp}



\end{mathpar}
\caption{Typing judgements.}\label{fig:typesystem}	
\end{small}
\end{figure*}

\subsubsection{Additional restrictions}\label{sec:restric}
In addition to the constraints imposed by our type system, some
operations require their operands to fulfill certain criteria in order
to be well-defined:
\begin{itemize}
  \item $\reduce[z]{f}{A}$ and $\reducebykey{f}{A}$: $f$ is associative
    and commutative, and $z$ is a neutral element for $f$.
  \item $\fixpoint[\delta]{R}{\varphi}$: $\varphi$ is a monoid
    homomorphism, $\delta$ is an aggregation function, and they are compatible.
\end{itemize}
The user needs to provide terms that satisfy these criteria since they
cannot be verified statically in general. 
However, regarding the homomorphism criterion for $\phi$, even though we cannot check
statically whether an arbitrary function is a monoid homomorphism, we can
identify a subset of functions that can be
statically checked. It is the set of terms $\varphi$ of the form
$\closure{X}{\mucriteria(X)}$ where $\mucriteria(X)$ is defined as
follows:
\begin{small}
  \smallsyntax{      
  \entry \mucriteria(X) [] 
        X []
      \oris \flatmapf{f}{\mucriteria(X)} [$X$ does not appear in $f$]
      \oris \join{\mucriteria(X)}{A}  [$X$ does not appear in $A$] 
      \oris \join{A}{\mucriteria(X)}  [$X$ does not appear in $A$]
  }
\end{small}%
This set of terms is in fact quite general: indeed, we know from
algebra that homomorphisms from $\tbag\valtyp$ to $\tbag\valtyp$ are
in one-to-one correspondance with functions from $\valtyp$ to
$\tbag\valtyp$, \emph{via} the general flatmap operation\footnote{This
  is due to the universal property of $\tbag\valtyp$, which is the
  free commutative monoid on $\valtyp$.}. In our case, flatmap has a
restriction relative to distributed bags, which is why we also have
$\opname{join}$; so the only monoid homomorphisms which cannot be
written in the form $\closure{X}{\mucriteria(X)}$ are functions which
manipulate distributed bags in a way which cannot be expressed as a
join. Thus, it makes sense to check statically whether the term
provided by the programmer is of that form and issue a warning if it
is not. 

For example, the program shown in Figure~\ref{fig:sppgm} can be expressed as a fixpoint $\fixpoint[\delta]{R}{\varphi}$ where $\varphi$ is an homomorphism of the form $\closure{X}{\mucriteria(X)}$ which is in charge of computing new paths and their lengths. We will further detail this example after having described the denotational semantics of the algebraic operators involved.

\packedsubsection{\mumonoid{} denotational semantics}\label{sec:densem}

Figure~\ref{fig:denotsem} gives the denotational semantics of the main
algebraic operations. It assumes all terms are well-typed \emph{and}
satisfy the additional restrictions mentioned in
Sec.~\ref{sec:restric}. Each closed term has a denotation in the
domain corresponding to its type, with the additional possible
denotation $\omega$ which belongs to all types and represents a
computation which does not terminate. Any of those operations returns
$\omega$ when applied to $\omega$.
\begin{figure*}[t]
  \begin{small}
      \begin{mathpar}
        \flatmapf{f}{A} = \biguplus_{a\in A} f(a)
        
        \reduce{\otimes}{A} = \bigotimes_{a\in A} f(a)
        
        \reducebykey{\otimes}{A} = \bagcomp{(k, \bigotimes_{(k, v)\in A}v)}{k\in \opname{keys}(A)}

        \cogroup{A}{B} = \bagcomp{(k,(\bagcomp{v}{(k,v)\in A}, \bagcomp{w}{(k,w)\in B}))}{k \in \opname{keys}(A)\sunion \opname{keys}(B)}
        
        \join{A}{B} = \bagcomp{(k, (v, w))}{(k, v)\in A \wedge (k, w)\in B}
        
        \fixpoint[\delta]{R}{\varphi} = \begin{cases}
        S_N &\text{if there exists } $N$ \text{ such that }
        S_{N+1} = S_N\\
        \omega &\text{otherwise}\end{cases} \text{ where } S_n =
\mathop{{\textstyle\bigotimes\nolimits_{\textstyle\delta}}}\limits_{0 \leq k \leq n}
        (\delta\circ\phi)^k(\delta(R))
      \end{mathpar}
  where 
     $keys(A) = \distinct(\setcomp{k}{(k,a) \in A})$; and
      $\sunion$ is distinct union of bags.
  \caption{Denotational semantics}\label{fig:denotsem}
  \end{small}
  \end{figure*}

These operations, except $\mu$, are monoid homomorphisms \cite{fegaras-jfp2017}, as
discussed in \ref{sec:homo}. We can check that they are still monoid
homomorphisms if we add $\omega$ to all the monoids as an absorbing
element\footnote{For a monoid $(S,\otimes,e)$, an absorbing element $\omega$ satisfies the following $\forall x\in S x\otimes \omega = \omega = \omega \times x$.}.

\paragraph{Properties of $\mu$}
  Recall that $\delta$ and $\phi$ being compatible means that
  $\delta\circ\phi$ is a monoid homomorphism: $\mondelta\to\mondelta$.  
  Thus we have: $$(\delta\circ\phi)(\mathop{{\textstyle\bigotimes\nolimits_{\textstyle\delta}}}\limits_{k\leq n}
        (\delta\circ\phi)^k(\delta(R))) =
        \mathop{{\textstyle\bigotimes\nolimits_{\textstyle\delta}}}\limits_{k\leq n}
        (\delta\circ\phi)^{k+1}(\delta(R)).$$
The only term missing to obtain $S_{n+1}$ on the right is
$(\delta\circ\phi)^0(\delta(R))$, i.\,e.\ $\delta(R)$. So we have, for
any $n$: $S_{n+1} = \delta(R)\opdelta(\delta\circ\phi(S_n))$. In other
words, if we use the definitions to `clean up' superfluous $\delta$s:
$S_{n+1} = \Psi(S_n)$ with $\Psi = X\mapsto \delta(R\bunion \phi(X))$.
So $S_{n+1}$ depends only on $S_n$, making the definition in
Fig.~\ref{fig:denotsem} consistent (if an $N$ is reached such that
$S_{N+1} = S_N$ then the sequence becomes stationary) and meaning
that $\fixpoint[\delta]{R}{\phi}$ is a
fixpoint\footnote{The least fixpoint, if we define an appropriate
  ordering relation on $\mondelta$, e.\,g.\ set inclusion in the
  standard case where $\delta = \distinct$.} of $\Psi$.

We now prove the following theorem, which is crucial for optimizing
distributed fixpoint computations:

\begin{thm}$\fixpoint[\delta]{\cdot}{\phi}$ is a monoid
homomorphism from $\tbag\valtyp\cup\{\omega\}$ to
$\mondelta\cup\{\omega\}$.
\end{thm}
\begin{proof}
$\fixpoint[\delta]{\emptybag}{\phi} = \emptybag$ is immediate.

Let $R_0 = R_1\bunion R_2$. For all $n$ and for $i$ in
  $0, 1, 2$, we write
  $S_n^i =
  \mathop{{\textstyle\bigotimes\nolimits_{\textstyle\delta}}}\limits_{k\leq
    n} (\delta\circ\phi)^k(\delta(R_i))$. Since $\delta$ is a
  homomorphism from bags to $\mondelta$ and $\delta\circ\phi$ is a
  homomorphism from $\mondelta$ to $\mondelta$, we have $S_n^0 =
  S_n^1\opdelta S_n^2$ for all $n$. Thus, if $(S_n^1)$ and $(S_n^2)$ both
  become stationary at some point, say $N_1$ and $N_2$, then $(S_n^0)$
  becomes stationary at $max(N_1, N_2)$ and we do have
  $\fixpoint[\delta]{R_0}{\phi} =
  \fixpoint[\delta]{R_1}{\phi}\opdelta\fixpoint[\delta]{R_2}{\phi}$. 
\end{proof}

\packedsubsection{Examples}\label{sec:examples}

We present in this section examples of recursive programs expressed in \mumonoid{}.

\paragraph{\textbf{Transitive closure} (TC)}
\begin{small}
  \begin{flalign*}
    &\kwlet~\textbf{reverse\_edges} = \closure{(a, b)}{\bag{(b, a)}}~\kwin\\
    &\kwlet~\textbf{drop\_mid} = \closure{(mid, (src, dest))}{\bag{(src, dest)}}~\kwin\\
  &\fixptn{R}{
    \closure{X}{\flatmapf{\textbf{drop\_mid}}{\join{\flatmapf{\textbf{reverse\_edges}}{X}}{R}}}}
  \end{flalign*}%
\end{small}%
  where $R$ is a dataset of tuples (source, destination) representing the edges of a graph.

  This expressions computes the entire transitive closure of the input graph $R$. %
    
  The sub-expression {\join{\flatmapf{\textbf{reverse\_edges}}{X}}{R}} joins a path from $X$ with a path from $R$ when the target node of the first path corresponds to the start node of the second path. This intermediary node is the join key, thus the join constructs pairs of the form $(mid, (src, dest))$ where $mid$ is the intermediary node, which must then be dropped by another flatmap.
  So, at each iteration, the paths in $X$ obtained in the last iteration get appended with edges from $R$ whenever possible. The computation ends when no new paths are found.
 
    \paragraph{\textbf{Shortest path} (SP)} 
    \begin{small}
    \begin{flalign*}
 &\kwlet~\textbf{key\_dest} = \closure{((src, dest), w)}{\bag{(dst, (src, w))}}~\kwin\\
 &\kwlet~\textbf{key\_src} = \closure{((src, dest), w)}{\bag{(src, (dest, w))}}~\kwin\\
 &\kwlet~\textbf{combine} = \closure{(mid, ((src, w_1), (dest, w_2)))}{\bag{((src, dest), w_1+w_2)}}~\kwin\\
 &\kwlet~\textbf{all\_paths} = \fixptn{R}{\closure{X}{\flatmapf{\textbf{combine}}
   {\join{\flatmapf{\textbf{key\_dest}}{X}}{\flatmapf{\textbf{key\_src}}{R}}}}}~\kwin\\
 &\reducebykey{\min}{\textbf{all\_paths}}
    \end{flalign*}%
    \end{small}%
    where $R$ is a dataset of tuples ((source, destination), weight) representing the weighted edges of a graph.


    The expression computes the shortest path between each pair of nodes in the input graph $R$. New paths are computed by performing a transitive closure while summing
     the lengths of the joined paths. Finally, the \reducebykeyop{} operation keeps the shortest paths between each pair of nodes.
     
     \paragraph{\textbf{Flights}}
     \newcommand\corr{\opname{corr}}
     \newcommand\Flight{\opname{Flight}}
     \newcommand\dtime{\opname{dtime}}
     \newcommand\atime{\opname{atime}}
     \newcommand\dep{\opname{dep}}
     \newcommand\dest{\opname{dest}}
     \newcommand\dur{\opname{dur}}
     \begin{small}
       \begin{flalign*}
       &\kwlet~\textbf{corr\_possible} = \closure{(\corr,(\Flight(\dtime1, \atime1, \dep1, \dest1, \dur1), \Flight(\dtime2, \atime2, \dep2, \dest2, \dur2)))}
       {\\&\qquad\qquad\ite{\atime1 < \dtime2}{\singleton{\Flight(\dtime1, \atime2, \dep1, \dest2, \dur1+\dur2)}}{\singleton{}}}~\kwin\\
       &\kwlet~\textbf{key\_dest} = \closure{\Flight(\dtime, \atime, \dep, \dest, \dur)}{\bag{(\dest, \Flight(\dtime, \atime, \dep, \dest, \dur))}}~\kwin\\
       &\kwlet~\textbf{key\_dep} = \closure{\Flight(\dtime, \atime, \dep, \dest, \dur)}{\bag{(\dep, \Flight(\dtime, \atime, \dep, \dest, \dur))}}~\kwin\\
       &\fixptn{R}{\closure{X}{\flatmapf{\textbf{corr\_possible}}{\join{\flatmapf{\textbf{key\_dest}}{X}}{\flatmapf{\textbf{key\_dep}}{R}}}}}
     \end{flalign*}%
     \end{small}%
          where $R$ is a dataset of direct flights. $\Flight(\dtime, \atime, \dep, \dest, \dur)$ is a flight object with a departure time $\dtime$, arrival time $\atime$, departure location $\dep$, destination $\dest$ and duration $\dur$.  %
           At each iteration, the fixpoint expression computes new flights by joining the flights obtained at the previous iteration with the flights dataset, in such a way that two flights produce a new flight if the first flight arrives before the second flight departs, and the first flight destination airport is the second's flight departure airport. The computation stops when no more new non-direct flights can be deduced.

     \paragraph{\textbf{Path planning}}
    \begin{small}
      \begin{flalign*}
        &\kwlet~\textbf{paths} = \closure{(\opname{City}(n_1,l_1),\opname{City}(n_2,l_2))}{\bag{(\opname{Path}(n_1,n_2),l_1 \lunion l_2)}}~\kwin\\
        &\kwlet~\textbf{key\_name\_dep} = \closure{(\opname{Path}(s,d),l)}{\bag{(s,(\opname{Path}(s,d),l))}}~\kwin\\
        &\kwlet~\textbf{key\_name\_dest} = \closure{(\opname{Path}(s,d),l)}{\bag{(d,(\opname{Path}(s,d),l))}}~\kwin\\
        &\kwlet~\textbf{combine} = \closure{(k,((\opname{Path}(s_1,d_1),l_1),(\opname{Path}(s_2,d_2),l_2)))}{\singleton{(\opname{Path}(s_1,d_2),l_1 \lunion l_2)}}~\kwin\\
        &\kwlet~\textbf{all\_paths} = \fixptn{\flatmapf{\textbf{paths}}{R}}{\closure{X}{\flatmapf{\textbf{combine}}{\join{\flatmapf{\textbf{key\_name\_dest}}{X}}{~\\
        &\qquad\qquad\flatmapf{\textbf{key\_name\_dep}}{\flatmapf{\textbf{paths}}{R}}}}}}~\kwin\\
        &\flatmap{(\opname{Path}(s,d),l)}{\ite{s = \text{"Paris" and } d = \text{"Geneva"}}{\singleton{(\opname{Path}(s,d),l)}}{\singleton{}}}{
            \\&\qquad\qquad\reducebykey{\opname{bestRated}}{\flatmap{(s,d,l)}{(\opname{Path}(s,d),l))}{\textbf{all\_paths}}}}    
    \end{flalign*}
    \end{small}%
    where R is a set of routes between two cities, represented as pairs of cities. Each city $\opname{City}(n,l)$ has a name $n$ and a list of landmarks $l$ and each landmark $\opname{Landmark}(n,r)$ has a rating $r$. 
    $\opname{bestRated}(l_1,l_2)$ is a function that returns the best set of landmarks based on its ratings.
    
    The fixpoint \textbf{all\_paths} computes the set of landmarks that can be visited for each possible path between each two cities. The final term then computes the best path between Paris and Geneva.

    \paragraph{\textbf{Movie Recommendations}}
    \begin{small}
      \begin{flalign*}
        &\kwlet~\textbf{users\_who\_like} = \closure{x}{\flatmap{\text{User}(u,bm)}{\ite{x \in bm}{bm}{\singleton{}}}{U}}~\kwin\\
      &\fixptn{S}{\closure{X}{\flatmapf{\textbf{users\_who\_like}}{X}}}
    \end{flalign*}%
    \end{small}%
where $U$ is a set of users, each user User$(u,bm)$ has a set of best movies $bm$.

The query computes a set of recommended movies by starting from a set of movies $S$ and by
adding the best movies of a user if one of his best movies is in the set of recommended movies until no new movie is added.

\pagebreak 
\packedsubsection{Evaluation of expressions}\label{sec:eval}

\nopagebreak

\packedsubsubsection{Local execution}\label{sec:locexec}

\nopagebreak

\paragraph{Pattern matching and function application}

\nopagebreak

The result of matching a value against a pattern is either a set of
pattern variable assignments or $\bot$. It is defined as follows:
\begin{align*}
  \patmatch\val\patvar &= \{\patvar\mapsto\val\}\\
  \patmatch{\caseclass{\val_1,\ldots,\val_n}}
  {\caseclass{\pat_1,\ldots,\pat_n}} &=
     \patmatch{\val_1}{\pat_1}\cup\cdots\cup\patmatch{\val_n}{\pat_n}\\
  \patmatch{\caseclass{\cdots}}{\caseclass[\classname']{\cdots}} &=
     \bot\text{ if } \classname\neq\classname'
\end{align*}
where we extend $\cup$ so that $\bot\cup S = \bot$.

A lambda expression
$f = \matchclosure{\pat_1\to\expr_1\mid\cdots\mid\pat_n\to\expr_n}$
contains a number of patterns together with return expressions. When
this lambda expression is applied on an argument $\val$ ($\app{f}{\val}$),
the argument is matched against the patterns in order, until the
result of the match is not $\bot$. Let $i$ be the smallest index such
that $\patmatch{\val}{\pat_i}=S\neq\bot$, the result of the application
is obtained by substituting the free pattern variables in $\expr_i$
according to the assignments in $S$.

\paragraph{Monoid homomorphisms}

The definition of algebraic operations as monoid homomorphism suggests
that they can be evaluated in the following way: if $\phi$ is a
homomorphism from $\tbag\valtyp$ to $(\valtyp', e, \otimes)$, $\phi(\bag{v_1,\ldots,
  v_n}) \evaluatesto{} \phi(\singleton{v_1}) \otimes \phi(\singleton{v_2}) \otimes...\otimes \phi(\singleton{v_n})$. 
As monoid operators are associative, parts of an expression in the form $e_1 \otimes e_2 \otimes ...\otimes e_n$ can be evaluated in any order and in parallel. 

\paragraph{Fixpoint operator}

The fixpoint operator can be evaluated as a loop, as described in Def.~\ref{def:mu}.
We can summarise it with the following reduction rules, where $\runmu$
represents a running $\mu$ computation and has the bag which
accumulates the results as an additional parameter:

\begin{small}
 \begin{mathpar}
\lbrack\initrule\rbrack~{\fixpoint[\delta]{R}{\phi}\evaluatesto{} \rfixpoint{\delta}{\delta(R)}{\phi}{\delta(R)}}

\lbrack\stoprule\rbrack~\inferrule{S\opdelta\varphi(R) = S} 
     {\rfixpoint{\delta}{R}{\phi}{S} \evaluatesto{} S}

\lbrack\looprule\rbrack~\inferrule{S\opdelta\varphi(R)\neq S}
     {\rfixpoint{\delta}{R}{\phi}{S} \evaluatesto{} \rfixpoint{\delta}{\delta(\phi(R))}{\phi}{S\opdelta\varphi(R)}}
 \end{mathpar}
\end{small}

\packedsubsubsection{Distributed execution}\label{sec:distexec}

We consider in a distributed setting that distributed bags are partitioned. Distributed data is noted in the following way: $R = R_1|R_2|...|R_p$, meaning that R is split into $p$ partitions stored on $p$ machines.
We can write a new slightly different version of the rule described above for evaluating partitioned data:
\begin{itemize}
\item $\phi(R_1|R_2|...|R_p) \evaluatesto{}
  \phi(R_1)|\phi(R_2)|...|\phi(R_p)$ if $\phi$ is an homomorphism from
  bags to bags (partitioning does not have to change)
\item
  $\phi(R_1|R_2|...|R_p) \evaluatesto{} \phi(R_1) \otimes^{nl}
  \phi(R_2) \otimes^{nl} ...\otimes^{nl} \phi(R_p)$ if $\phi$ is an
  homomorphism from bags to $(M, e, \otimes)$, where $\otimes^{nl}$ is
  the non-local version of $\otimes$. Applying this non-local
  operation means that data transfers are required.
\end{itemize}
    
This means that in our algebra, all operators apart from
\opname{flatmap}, or \opname{join} when one of the parameters is a
local bag, need to send data across the network (for executing the
non-local version of their monoid operator). The execution of these
non-local operators depends on the distributed platform. Spark for
example performs \emph{shuffling} to redistribute the data across
partitions for the computation of certain of its operations like
cogroup and groupByKey.


\packedsection{Optimizations} \label{sec:optim} 

In this section, we propose new optimization rules for terms with fixpoints, and  describe when and how they apply.  The purposes of the rules are (i) to identify which basic operations within an algebraic term can be rearranged and under which conditions, and (ii) to describe how new terms are produced or evaluated after transformation.

We first give the intuition behind each optimization rule before zooming on each of them to formally describe when they apply.  The four new optimization rules are:

\begin{packeditemize}
\item $\pushingfilterRule{}$ is a rewrite rule of the form: $$ \filtername{\fixptn{R}{\varphi}} \rewritesto{}\fixptn{\filtername{R}}{\varphi}$$ it aims at pushing a filter \filtername{} inside a fixpoint, whenever this is possible. A filter is a function which keeps only some elements of a dataset based on their values; we define it formally in Sec.~\ref{sec:filter}.

\item $\semijoinRule{}$ is a rewrite rule of the form: 
 $$\join{A}{\fixptn{R}{\varphi}} \rewritesto{} \join{A}{\fixptn{\semijoinfilter{R}}{\varphi}} $$
it aims at inserting a filter $\semijoinfilter{}$ inside a fixpoint before a join is performed. It is inspired by the semi-join found in relational databases, and tailored for \mumonoid{}.
\item $\pushingaggregRule$ is a rewrite rule of the form:
$$\delta(\fixptn{R}{\varphi})  \rewritesto{} \fixpoint[\delta]{R}{\varphi}$$
It aims at pushing an aggregation function $\delta$ inside a fixpoint,
transforming a simple fixpoint into a fixpoint with aggregation. This
rule requires $\delta$ to be compatible with $\varphi$;
it is inspired from the premappability condition in Datalog \cite{zaniolo-tplp17}.
\item an optimization rule $\fixpointdistRule{}$ that determines how a fixpoint term is evaluated in a distributed manner by choosing among two possible execution plans. 
\end{packeditemize}

\packedsubsection{Pushing filter inside a fixpoint ($\pushingfilterRule$)}
\packedsubsubsection{Filter depending on a single pattern variable}\label{sec:filter}
\begin{defi}[filter]\label{def:filter}
  We call \emph{filter} a function of the form: $$\closure{D}{\flatmap{\pi}{\ite{c(a)}{\singleton{\pi}}{\singleton{}}}{D}}$$ where $\pi$ is a pattern containing the variable $a$ and $c(a)$ is a Boolean condition depending on the value of $a$.

  Such a function returns the dataset $D$ filtered by retaining only the elements whose value for $a$ (as determined by pattern-matching that element with $\pi$) satisfies $c(a)$. The elements are unmodified, so the result is a subcollection of $D$.
\end{defi}

In the following, we consider a filter $\filtername{}$ with $\pi$ and $a$ defined as above, and we denote by $\pi_a$ the function that matches an element against $\pi$ and returns the value of $a$ ($\pi_a = \closure{\pi}{a}$). For instance, $\pi_a((1,(5,6))) = 5$ for $\pi=(x,(a,y))$.

Let us consider a dataset $D$. In terms of denotational semantics, with the notations above, we have $F(D) = \{d\in D \mid c(\pi_a(d))\}$.

\paragraph{The $\pushingfilterRule$ rule}\label{sec:pushingfilter}
This rule consists in transforming an expression of the form $F(\fixptn{R}{\varphi})$ to an expression of the form $\fixptn{F(R)}{\varphi}$, where F is a filter. 

In the second form, the filter is pushed before the fixpoint operation. In other words, the constant part R is filtered first before applying the fixpoint on it. 
We now present sufficient conditions for the two terms to be equivalent.

\paragraph{$\pushingfilterRule$ condition}
Let (C) be the following condition: $$\forall r \in R  \quad \forall s \in \varphi(\singleton{r}) \quad \pi_a(r) = \pi_a(s)$$
Intuitively, this condition means that the operation $\varphi$ does not change the part of its input data that corresponds to $a$ in the pattern $\pi$, which is the part used in the filter; so for each record in the fixpoint that does not pass the filter, the record in R that has originated it does not pass the filter and the other way round. That is why we can just filter R in the first place.

Let $A =
  \flatmap{\pi}{\ite{c(a)}{\singleton{\pi}}{\singleton{}}}{\fixptn{R}{\varphi}}$.
We prove that if (C) is satisfied, then $A = \fixptn{F(R)}{\varphi}$.
To prove this, we use the following property of fixpoints
where $\delta=\distinct$:
\begin{lem}\label{lem:origin}
  $\forall a \in \fixptn{R}{\varphi} \;\;\; a \in \varphi^{(n)}(\singleton{r})$ for some $r \in R$ and $n \in N$
\end{lem}
\begin{proof}
We have: $$\fixptn{R}{\varphi} = \bigcup_{n \in \N} \varphi^{(n)}(R) = \bigcup_{n \in \N} \varphi{(\biguplus_{r \in R}\singleton{r})} = \bigcup_{n \in \N} (\biguplus_{r \in R}\varphi(\singleton{r})) = \bigcup_{n \in \N} (\bigcup_{r \in R}\varphi(\singleton{r}))$$
\end{proof}
Using the above lemma and condition (C), we have:
$$(*) \qquad\qquad \forall s\in \fixptn{R}{
  \varphi} \quad \exists r \in R \quad \pi_a(r) = \pi_a(s)$$

We now prove $A = \fixptn{F(R)}{\varphi}$  by proving the two inclusions:
\begin{enumerate}

\item $\fixptn{F(R)}{\varphi} \subset A$:

$F(R) \subset R \implies \fixptn{F(R)}{\varphi} \subset \fixptn{R}{\varphi} \quad$ (because $\fixptn{R}{\varphi} = \fixptn{F(R) \bunion R'}{\varphi} = \fixptn{F(R)}{\varphi} \sunion \fixptn{R'}{\varphi}$ and \fixptn{R}{\varphi} does not contain duplicates)

Let $s \in \fixptn{F(R)}{\varphi}$

$\exists r \in F(R) \quad \pi_a(s) = \pi_a(r)  \quad (*)$

So $c(\pi_a(s)) = c(\pi_a(r))=\text{true \quad (because } r\in F(R))$

So $s\in \fixptn{R}{\varphi}$ and $c(\pi_a(s))$ is true, then $s\in A$

\item $A \subset \fixptn{F(R)}{\varphi} $: 

Let $s\in A. \text{ We have:} \quad s \in \fixptn{R}{\varphi} \text{ and } c(\pi_a(s)) = \text{true}$ 

So $\exists r \in R \quad \exists n \in \N \quad \pi_a(s) = \pi_a(r) \quad (\text{because } (*) \text{ and } s \in \varphi^{(n)}(\singleton{r}))$

So $c(\pi_a(r)) = c(\pi_a(s)) = \text{true}$ 

So $r \in F(R)$, which means $\distinct(\varphi^{(n)}(\singleton{r})) \subset \fixptn{F(R)}{\varphi}$, so $s \in \fixptn{F(R)}{\varphi}$.

\end{enumerate}

\paragraph{Verifying the condition (C) using type inference}
We will start by explaining the intuition behind this before going into the details. 

For the condition (C) to hold, we need to make sure that the part of the data extracted by $\pi_a$ is not modified by $\varphi$.
For this, our solution is inspired by the idea that the type of a
parametric polymorphic function tells us information about its
behaviour \cite{theoremsforfree}: for example, if $f$
is a polymorphic function whose argument contains exactly one value of
the undetermined type $\alpha$ and whose result must also contain a
value of type $\alpha$, then the $\alpha$ value in the result is
necessarily the one in the argument ($f: \alpha \to \alpha \implies
\forall x \; f(x) = x$).

This reasoning can also be used for a more complex input type
$C(\alpha)$ that contains a polymorphic type $\alpha$. For instance:
$C(\alpha) = A(B(\alpha),D)$ is such a type given that $A$,$B$ and $D$
are type constuctors. So our goal is, given that $\varphi$ takes as
input a bag of elements of type C, to find an appropriate polymorphic
type $C(\alpha)$ that will be used for type checking $\varphi$. In
practice, we translate the $\varphi$ operation to a Scala function
that takes a polymorphic input type and use the Scala type inference
system \cite{Odersky04anoverview} to get the output type\footnote{We
  consider it a more practical solution than implementing our own type
  inference system supporting polymorphism.}. $C(\alpha)$ should be
built in such a way that the position of $\alpha$ in $C(\alpha)$ is
the same as the position of $a$ in $\pi$. Such a type is possible to
build because the type $C$ matches the pattern $\pi$, otherwise the
filtered term would not be type correct. Finally, if the output type also contains
the type $\alpha$ and has the same position as $a$ in $\pi$ then we
can show that the condition (C) holds. Note that we do not need a
full-fledged parametricity theorem for this: we only use the fact that
the Scala type system has singleton types for all values.

\paragraph{Building $C(\alpha)$}
Types are made from type constructors and basic types, and patterns are made from type constructors and pattern variables. So we can represent their structures using trees. In the following we sometimes refer to types by the trees representing them.

\begin{defi}[path]
We define the \emph{path} to the node labelled $n$ in the tree $T$ denoted path$(n,T)$ by the ordered sequence Seq($a_i$) where $a_i$ is the next child arity of the $i$th visited node to reach $n$ from the root of the tree. A node in a tree can be identified by its path.
\end{defi}

Let us consider the function $\texttt{replace}_\alpha(p, T)$ that, given a path $p$ and a type $T$ returns a polymorphic type $T(\alpha)$ that is obtained by replacing in $T$ the node at path $p$ and its children by a node labelled $\alpha$. 
Let us now consider $C(\alpha) = replace_\alpha(path(a,\pi), C)$, where $\tbag{C}$ is the input type of $\varphi$. Note that this path makes sense in $C$ because $C$ matches $\pi$ (see Appendix~\ref{sec:typesys}).

With $C(\alpha)$ built this way, we have the following: 
\begin{flalign}
&e:C \text{ and } \pi_a(e):\alpha \implies e:C(\alpha) \\
&e:C(\alpha) \implies \pi_a(e):\alpha 
\end{flalign}

For example:

\begin{small}
  \begin{center}
\begin{tabular}{ccc}
\qquad\Tree[.Tuple [.a ]  [.b ] [.c ]] \qquad&\qquad\Tree[.Tuple [.Tuple [.Int ]  [.Int ]] [.String ] [.Int ]] \qquad&\qquad\Tree[.Tuple [.$\alpha$ ] [.String ] [.Int ]]\qquad\\

\qquad$\pi$ &\qquad$C$ &\qquad$C(\alpha)$
\end{tabular}%
\end{center}
\end{small}%

We show that if $\varphi: \tbag{C(\alpha)} \to \tbag{C(\alpha)}$ then the condition (C) is verified:

Let $r \in R$ and let us take $\alpha = \singleton{\pi_a(r)}$ which is the singleton type containing the value $\pi_a(r)$. Since $\pi_a(r):\alpha$ and $r:C$, we have $r:C(\alpha)$ according to (1). 

We also have $\varphi(\singleton{r}):\tbag{C(\alpha)}$ because $\singleton{r}:\tbag{C(\alpha)}$ and $\varphi: \tbag{C(\alpha)} \to \tbag{C(\alpha)}$ which means that $\forall s  \in \varphi(\singleton{r}) \quad s: C(\alpha)$. So $\pi_a(s):\alpha$ (according to (2)) so $\pi_a(s) = \pi_a(r)$ hence (C).

\packedsubsubsection{Filters depending on multiple variables}
We showed that (C): $\forall r \in R \quad \forall s \in \varphi(\singleton{r}) \quad \pi_a(r) = \pi_a(s)$ is sufficient for pushing the filter in a fixpoint when the filter condition depends on $a$. %
We can easily show that when the filter depends on a set of pattern variables $V$, the sufficient condition becomes: $\forall r \in R \quad \forall s \in \varphi(\singleton{r}) \quad \forall v \in V \quad \pi_v(r) = \pi_v(s)$. So, if one of the variables in $V$ does not satisfy the condition the filter would not be pushed. However, we can do better by trying to split the condition $c$ to two conditions $c_1$ and $c_2$, such that $c = c_1 \land c_2$ and $c_1$ depends only on the subset of variables that satisfies the condition (this splitting technique is used in \cite{FegarasN18} to push filters in a \opname{cogroup} or a \opname{groupby}). If such a split is found, the filter \flatmap{\pi}{\ite{c}{\singleton{\pi}}{\singleton{}}}{R} can be rewritten as 
\flatmap{\pi}{\ite{c_2}{\singleton{\pi}}{\singleton{}}}{\flatmap{\pi}{\ite{c_1}{\singleton{\pi}}{\singleton{}}}{R}}. The inner filter can then be pushed.

\packedsubsection{Filtering inside a fixpoint before a join ($\semijoinRule{}$)}\label{sec:joinoptim}
Let us consider the expression: \join{A}{B}, where $B = \fixptn{R}{\varphi}$. After the execution of the fixpoint, the result is going to be joined with $A$, so only elements of this result sharing the same keys with $A$ are going to be kept. So in order to optimize this term, we want to push a filter that keeps only the elements having a key in $A$. This way, elements not sharing keys with $A$ are going to be removed before applying the fixpoint operation on them.

\begin{enumerate}
\item  we show that $join(A,B) = join(A, \semijoinfilter{B})$, where
  $\semijoinfilter{B} = \setcomp{(k,v)}{(k,v) \in B \wedge \exists w \;
    (k,w) \in A} $:

We have $join(A,B) = \setcomp{(k,(x,y))}{(k,x) \in A \wedge (k,y) \in B}$.

So $(k,(x,y)) \in join(A,B) \iff (k,x) \in A \land (k,y) \in B \iff (k,x) \in A \land ((k,y) \in B \land \exists w \; (k,w)~\in~A) \iff (k,x) \in A \;\land\; (k,y) \in \semijoinfilter{B} \iff (k,(x,y)) \in join(A,\semijoinfilter{B}))$.
\item  we show that $\semijoinfilter{B}$ is a filter on B. This filter
  can be pushed when the criteria on pushing filters is fulfilled:

We can show that
$\semijoinfilter{B} =
\flatmap{(k,v)}{\ite{c(k)}{\singleton{(k,v)}}{\singleton{}}}{B}$ where $c(k)$ is the boolean expression that corresponds to the predicate $\exists w (k,w) \in A$.
This expression can be: $c(k) = \reduce{\vee}{\flatmap{(k',a)}{k==k'}{A}}$.
Which means that in case $\varphi$ fulfills the criteria for pushing filters we will have $\semijoinfilter{B} = \semijoinfilter{\fixptn{R}{\varphi}} = \fixptn{\semijoinfilter{R}}{\varphi}$.

\item  we show as well that $\semijoinfilter{B} = C$
where:
$$C = \flatmap{(k, (s_x, s_y))}{ \ite{s_y \neq \singleton{}}{\flatmap{x}{\singleton{(k,x)}}{s_x}}{\singleton{}}}{\cogroup{B}{A}}$$

We have: $C =\biguplus_{(k,(s_x, s_y)) \in \cogroup{B}{A}}\biguplus_{x \in s_x} (\ite{s_y \neq \singleton{}}{\singleton{(k,(x,y))}}{\singleton{}})$

(1) Let $e \in C$.
So $\exists (k,(s_x, s_y))) \in \cogroup{B}{A}$ such that $\exists x\in s_x \quad e = (k,x)$ and $s_y \neq \singleton{}$.

We have $(k,(s_x, s_y)) \in \cogroup{B}{A}$, so $s_x = \setcomp{v}{(k,v) \in B}$, which means that $(k,x) \in B$ because $x \in s_x$.
And $s_y \neq \singleton{}$ means that $\exists w \; (k,w) \in A$,
so $e = (k,x) \in \semijoinfilter{B}$.

(2) Let $(k,x) \in \semijoinfilter{B}$.
We have $(k,x) \in B$ and $\exists w \; (k,w) \in A$.
So $k \in \opname{keys}(A) \sunion \opname{keys}(B)$.

Let $s_x = \setcomp{v}{(k,v) \in B}$ and $s_y = \setcomp{v}{(k,y) \in A}$, so  $(k, (s_x, s_y)) \in \cogroup{B}{A}$.

Since $x \in s_x$ and  $s_y \neq \singleton{}$ (because $\exists w \; (k,w) \in A$), then $(k,x) \in C$.
\end{enumerate}

\subsection{Pushing aggregation into a fixpoint ($\pushingaggregRule$)}\label{sec:pushagg}

\paragraph{The $\pushingaggregRule$ rule}
 consists in rewriting a term of the form
 $\delta(\fixptn{R}{\varphi})$ to a term of the form
 $\fixpoint[\delta]{R}{\varphi}$. It requires that $\delta$ is an
 aggregation function and compatible with $\phi$.

It is correct thanks to the following lemma:
\begin{lem}
  Let $\phi$ be a monoid homomorphism: $\tbag\valtyp\to\tbag\valtyp$,
  $\delta$ an aggregation function: $\tbag\valtyp\to\tbag\valtyp$
  compatible with $\phi$, and $R: \tbag\valtyp$ a dataset. Assume
  $\fixpoint{R}{\varphi}\neq\omega$ (i.\, e.\ the computation
  terminates). Then $\delta(\fixptn{R}{\varphi}) =
  \fixpoint[\delta]{R}{\varphi}$.
\end{lem}
\begin{proof}
  Let $(S_n)$ and $(S'_n)$ be the $S$ sequences corresponding
  respectively to the two fixpoints; thus we have $S_{n+1} = R\bunion
  \phi(S_n)$ and $S'_{n+1} = R\opdelta\phi(S'_n) = \delta(R\bunion
  \phi(S'_n))$. We prove by induction on $n$ that
  $\delta(S_n) = S'_n$ for any $n$: for $n = 0$ we have $\delta(S_0) =
  \delta(R) = S'_0$. Assume $S'_n = \delta(S_n)$, we have:
    \begin{align*}
      \delta(S_{n+1}) &= \delta(R\bunion\phi(S_n)) = \delta(\delta(R)\bunion\delta(\phi(S_n))) &(\delta \text{ is an
  aggregation function})\\
&= \delta(\delta(R)\bunion\delta(\phi(\delta(S_n))) &(\delta\circ\phi
= \delta\circ\phi\circ\delta)\\
&= \delta(\delta(R)\bunion\delta(\phi(S_n'))) &(\text{induction
  hypothesis})\\
&= \delta(R\bunion\phi(S_n')) = S'_{n+1}&(\delta \text{ is an aggregation
  function})
    \end{align*}
Then the result propagates to the fixpoint since we assumed that $\fixpoint{R}{\varphi}\neq\omega$.
\end{proof}
Applying this 
optimization on the expression of the SP example (Sec.\ref{sec:examples}) means that only 
the shortest paths are kept at each iteration of the fixpoint 
so we avoid computing all possible paths before keeping only the shortest ones
at the end.
The requirement of compatibility of $\delta$ with $\varphi$ means that the application of $\delta$ first before the $\varphi$ operation does not impact the result compared to when it is
applied once at the end. For instance, if we are computing the shortest paths between $a$ and $b$, we look for all paths between $a$ 
and $c$, append them to paths from $c$ to $b$, then keep the shortest ones. Alternatively, we could start by keeping only 
the shortest paths between $a$ and $c$ then append them to paths
between $c$ and $b$ without altering results.
At present, we do not have a method for statically checking this constraint. 
So, in practice, we require an annotation from the programmer on the
aggregation operations that verify the necessary constraints. However,
we can list common known aggregation functions :
$\reducebykey{f}{\cdot}$, filters (see Def.~\ref{def:filter}), mainly.

\packedsubsection{Distribution of the fixpoint operations ($\fixpointdistRule$)}\label{sec:distfixpoint}
As explained in \ref{sec:locexec}, the fixpoint operation is computed
locally using a loop (defined by $\initrule, \looprule$ and $\stoprule$). To
evaluate the fixpoint in a distributed setting, we could simply write
a loop that distributes the computation of the operation that is
performed at each iteration ($S\opdelta\phi(R)$) among
the workers. We call this execution plan $\plangloballoop$.
$\plangloballoop$ performs $\delta$ at each iteration on the whole
intermediary distributed bag $S$ to compute $S\opdelta\phi(R)$, which
in most cases (i.\, e.\ unless $\delta$ is the identity function)
requires synchronisation and data transfer between workers at each iteration. In the TC example (Sec.~\ref{sec:examples}), this plan amounts to appending, at each iteration, all currently found paths from all partitions with the graph edges R.

Alternatively, if we use the fact that \fixpoint[\delta]{R}{\varphi}
is a monoid homomorphism, then we can replace $\initrule$ with the
following distributed version (recall that $R_1|R_2|...$ denotes a
distributed bag split across different partitions $R_i$.
$\opdelta^{nl}$ denotes the non-local version of $\opdelta$):
$$\fixpoint[\delta]{R_1|R_2|...}{\varphi} \evaluatesto \rfixpoint{\delta}{\delta(R_1)}{\phi}{\delta(R_1)}\opdelta^{nl}\rfixpoint{\delta}{\delta(R_2)}{\phi}{\delta(R_2)}\opdelta^{nl}...$$

Then each $\rfixpoint{\delta}{\delta(R_i)}{\phi}{\delta(R_i)}$ is going to be evaluated by
$\looprule$ and $\stoprule$ as they are fixpoints on local bags. This
execution plan, that we name $\planparallelloop$, will avoid doing
non-local set unions or aggregations between all partitions at each
iteration of the fixpoint. Instead, the fixpoint is executed locally
on each partition on a part of the input, after which the aggregate
$\opdelta^{nl}$ is computed once to gather results. In our example,
this amounts to computing, on each partition $i$, all paths in the
graph starting from nodes in $R_i$; the result is then the union of
all obtained paths.

This reduction in data transfers can lead to a significant improvement of performance, since the size of data transfers over the network is a determining factor of the performance of distributed applications.

The optimization rule $\fixpointdistRule$ uses the plan $\planparallelloop$ instead of $\plangloballoop$ for evaluating fixpoints.

\packedsubsubsection{Avoiding $\sunion^{nl}$ in $\planparallelloop$}\label{sec:distfppartitioning}
In the common case where $\delta$ is $\distinct$, $\planparallelloop$ can be optimized further by repartitioning the data in the cluster in such a way that every result of the fixpoint appears in one partition only. When that is the case, it is sufficient to perform a bag union rather than a set union that removes duplicates from across the cluster. If we know that there is a part in the input that does not get modified by $\varphi$, we can repartition the data on this part of the input (no two different partitions have the same value for this part), so the result of the fixpoint is also going to be repartitioned in the same way. We formalize this optimization in the following way:

Let $\pi$ a pattern that matches the input of $\varphi$ and $a$ a pattern variable in $\pi$. We consider the following propositions:
\begin{flalign*}
&(C_a): \forall r \in R  \quad \forall s \in \varphi(\singleton{r}) \quad \pi_a(r) = \pi_a(s)\\
 &(P_a): \forall i \neq j \;\forall x \in R_i\; \forall y \in R_j\quad \pi_a(x) \neq \pi_a(y)
\end{flalign*}
\begin{lem} If there exists a pattern variable $a$ that verifies
  $(C_a)$, then: $$P_a \implies \forall i \neq j
  \quad\fixptn{R_i}{\varphi} \cap  \fixptn{R_j}{\varphi} =
  \emptyset$$\end{lem}
\begin{proof}
Let us suppose there exist a pattern variable $a$ for which $(C_a)$ is verified, and let us suppose $(P_a)$.
Let $R_i$ and $R_j$ partitions of $R$ such that $i \neq j$.
$(C_a)$ implies $\forall s\in \fixptn{R}{ \varphi} \quad \exists r \in
R \quad \pi_a(r) = \pi_a(s)$ because of Lemma~\ref{lem:origin}.
Which means that for any $x \in \fixptn{R_i}{\varphi}$ and $y \in \fixptn{R_j}{\varphi}$, $\exists r_i \in R_i \; \exists r_j \in R_j \;\;\; \pi_a(r_i) = \pi_a(x) \text{ and } \pi_a(r_j) = \pi_a(y)$. We have $\pi_a(r_i) \neq \pi_a(r_j)$ because $(P_a)$, so $x \neq y$. Hence $\forall i \neq j \;\; \fixptn{R_i}{\varphi} \cap \fixptn{R_j}{\varphi} = \emptyset$.
\end{proof}
This means that $ \fixptn{R_1 \sunion \varphi(R_1)}{\varphi} \; \sunion^{nl} \; \fixptn{R_2 \sunion \varphi(R_2)}{\varphi} \; \sunion^{nl}... = \fixptn{R_1 \sunion \varphi(R_1)}{\varphi} \; | \; \fixptn{R_2 \sunion \varphi(R_2)}{\varphi} \; |...$

The pattern variable $a$ that verifies $(C_a)$ can be found by using the technique explained in Section~\ref{sec:filter}. We explore every node $n$ in $C$ (\tbag{C} is the input type of $\varphi$) starting from the root of $C$ and we build $C(\alpha) = \opname{replace}_\alpha(\opname{path}(n,C),C)$ until we find a node that verifies $\varphi:C(\alpha) \to C(\alpha)$.

If such $a$ is found, we repartition the data according to $(P_a)$ by using the API provided by the big data platform on which the code is executed, given that $a$ can be extracted from the input data using pattern matching.

\packedsubsection{Effects of the rules on performance}~\label{sec:ruleseffects}

In this section, we discuss the impacts of the rules and the conditions under which they produce terms that are more efficient in practice.
The verification of these conditions is outside of the scope of this paper.
Techniques that estimate the size of algebraic expressions such as those found in~\cite{lawal-cikm20} can be used to perform such verifications.

\packedsubsubsection{\pushingfilterRule{} effects}
Rule $\pushingfilterRule$ is a logical optimization rule in the sense that the term it produces is always more efficient than the initial term. Indeed, a filter reduces the size of intermediate data. The application of $\pushingfilterRule$ thus reduces data transfers. Operators are also executed faster on smaller data. The application of $\pushingfilterRule$ can thus only improve performance. 

\packedsubsubsection{$\semijoinRule$ effects}

The rule $\semijoinRule$ introduces an additional \opname{cogroup} to compute the filter being pushed in the fixpoint (as detailed in Sec.~\ref{sec:joinoptim}).
The cost of evaluating a term depends on two important aspects: the size of non-local data transfers it generates, and the local complexity of the term (i.e. the time needed for executing its local operations). %

$\semijoinRule$ can improve local complexity. The reason is that the additional \opname{cogroup} is evaluated only once, whereas the pushed filter makes $R$ (the first argument of the fixpoint \fixpoint{R}{\varphi}) smaller. %
Therefore, in general, each iteration of the fixpoint is executed faster as it deals with increasingly less data (each value removed from the initial bag would have generated more additional values with each iteration). The final join with the result of the fixpoint also executes faster because its size is reduced prior to the join.
In the worst case (the filter does not remove any result), the additional cogroup does not change the worst case complexity of the computation (\opname{join} and \opname{cogroup} have the same worst case complexity $O(n^2)$).

To analyse the impact of the rule on non-local data transfers, we need to estimate and compare the size of the transfers incurred by the terms: $join(A, \fixptn{R}{\varphi})$ and $join(A, \fixptn{\semijoinfilter{R}}{\varphi})$ (obtained after applying the rule). As mentioned in Section~\ref{sec:distexec}, all our algebraic operators apart from \opname{flatmap} trigger non-local transfers. 
Let $size(t)$ be the size of the result obtained by evaluating $t$, $size_t(t)$ the size of transfers incurred by the evaluation of $t$, and $N$ the number of partitions (parallel tasks) in the cluster.  
We consider the following:

\begin{itemize}
  \item Repartitioning a dataset $A$ by key requires all $A$ to be tranferred across the network (each element of $A$ has to be in the partition corresponding to its key). This means that $\sizetran{\groupby{t}} = size(t)$, and $\sizetran{\cogroup{t_1}{t_2}} = \sizetran{\join{t_1}{t_2}} = \size{t_1} + \size{t_2}$.
  \item $\sizetran{\distinct(A)} = N \times \size{A}$ because all $A$ has to be seen by each partition so that duplicates can be removed globally. This means that $\sizetran{\fixptn{R}{\varphi}} = N \times \size{\fixptn{R}{\varphi}}$.
\end{itemize}

Let $S_1 = \sizetran{\join{A}{\fixptn{R}{\varphi}}}$ and $S_2 = \sizetran{\join{A}{\fixptn{\semijoinfilter{R}}{\varphi}}}$.
So we have: 

$S_1 = \size{A} + (N+1) \times \size{\fixptn{R}{\varphi}}$, here the result of the fixpoint is sent twice: the first time to compute the fixpoint and the second time to compute the join between A and the fixpoint result.

$S_2 = 2 \times \size{A} + (N+1) \times \size{\fixptn{\semijoinfilter{R}}{\varphi}} + \size{R}$, here $\semijoinfilter{R}$ requires making a \opname{cogroup} between $A$ and $R$ which incurs an additional transfer of their sizes. On the other hand, only a filtered fixpoint result is sent.

In order to determine if $\semijoinRule$ improves data transfers we need to compare $S_1$ and $S_2$, which amounts to comparing the following quantities: $(N+1) \times \size{\fixptn{R}{\varphi}}$ and $(N+1) \times \size{\fixptn{\semijoinfilter{R}}{\varphi}} + \size{A} + \size{R}$. 
In other words, $\semijoinRule$ improves data transfers when the data removed from the fixpoint result (by pushing the filter into it) makes up for the sizes of $A$ and $R$ that are transferred to compute the additional \opname{cogroup}. 
This is likely to be the case since the data obtained at each iteration of the fixpoint (including R) is filtered.

\packedsubsubsection{$\pushingaggregRule$ effects}\label{sec:pushaggcriteria}

The $\pushingaggregRule$ rule applies the aggregation function $\delta$ on the
fixpoint’s intermediate results instead of once at the end. Whenever
$\delta$ reduces the size of these results, the fixpoint operation deals with less data at each iteration (which also generally reduces the number of iterations).
For example, if we are computing the shortest paths, applying the rule would mean that we are only going to deal with the shortest paths at each step instead of the entirety of possible paths. This can also lead to the termination of the program in case the graph has cycles (note that the programs are semanticaly equivalent but the evaluation of the first does not terminate). 
Additionally, when $\fixpointdistRule$ is applied, $\pushingaggregRule$ can only reduce the size of the data transferred across the network because $\delta$ is executed locally and reduces the sizes of the local fixpoints.

\packedsubsubsection{$\fixpointdistRule$ effects}

Application of $\fixpointdistRule$ can drastically decrease data transfers across the network. As explained in Sec.~\ref{sec:distfixpoint}, plan $\planparallelloop$ avoids transferring intermediate results during fixpoint iterations or even entirely (if a data partioning that verifies the criteria presented in ~\ref{sec:distfppartitioning} exists).

The efficiency of the two plans that distribute the fixpoint depends on two aspects. %
First, for a term $\fixptn{R}{\varphi}$ to be evaluated on a plateform like Spark, the collections referenced in $\varphi$ have to be available locally in each worker so that it can compute the fixpoint locally. For instance, if $\varphi = join(X,S)$ then $S$ and $X$ (at each iteration) are both referenced by $\varphi$. This is a limitation of plan $\planparallelloop$: when those datasets become too large to be handled by one worker, $\plangloballoop$ is more appropriate. %
Second, a factor that determines the efficiency of $\planparallelloop$ (and impacts the size of the  iteration results $X$) is the number of parallel tasks that execute the program. In Spark, this corresponds to the number of partitions.
 Increasing the number of partitions increases the parallelization and reduces the load on each worker because the local fixpoints start from smaller constant parts. For a term $\fixptn{R}{\varphi}$, it is thus possible to regulate the load on the workers by splitting $R$ into smaller $R_i$, resulting in smaller tasks on more partitions. 
The ideal number of partitions is the smallest one that makes all workers busy for the same time period, and for which the size of the task remains suitable for the capacity of each worker. Increasing the number of partitions further would only increase the overhead of scheduling. %
Thus, estimation of an appropriate number of partitions for $\planparallelloop$ would ideally be based on an estimated size of the constant part, the size of intermediate data produced by the fixpoint and the workers memory capacity. In the experiments we present below, we use a simple heuristic to determine the number of partitions: 4 times the total number of cores of the cluster.

\packedsection{Experimental results} \label{sec:experiments}
\paragraph{Methodology.}
We experiment the \mumonoid{} approach in the context of the Spark platform \cite{ZahariaXWDADMRV16}. 

We evaluate Spark programs generated from optimized \mumonoid{} expressions, and compare their performance with the state-of-the-art implementations Emma\cite{alexandrovKKM16} and DIQL \cite{FegarasN18}, which are Domain Specific Languages (more detail about them in Section~\ref{sec:relatedworks}). The authors of Emma showed that their approach outperforms earlier works in \cite{alexandrovKKM16}. DIQL is a DSL built on monoid algebra (of which the \mumonoid{} algebra is an extension). Comparing against DIQL shows the interest of having a first-class fixpoint operator in the monoid algebra.
\paragraph{Experimental setup.}
Experiments have been conducted on a Spark cluster composed of 5 machines (hence using 5 workers, one on each machine, 
and the driver on one of them). Each machine has 128~Go of RAM and the Spark worker on this machine is configured to use 40~GB, 2 Intel Xeon E5-2630 v4 CPUs (2.20 GHz, 20 cores each) 
and 66~TB of 7200 RPM hard disk drives, running Spark 2.2.3 and Hadoop 2.8.4 inside Debian-based Docker containers.

\paragraph{Algorithms}
The algorithms considered in these experiments are: TC, SP, Flights, Path Planning, and Movie Recommendations presented in the examples (Section~\ref{sec:examples}). 
In addition, we evaluate two variants of TC and SP: TC filter and SP filter, 
where we compute the paths starting from a subset of 2000 nodes randomly chosen in the input graph. 

\paragraph{Systems}
\mumonoid{} is evaluated against other systems on the algorithms mentioned above. 
\mumonoid{} programs are generated from the \mumonoid{} terms expressing these algorithms (see Examples Section~\ref{sec:examples}) and by systematically applying the rules $\pushingfilterRule, \semijoinRule, \pushingaggregRule, \fixpointdistRule$ (of Section~\ref{sec:optim}).
We evaluate these programs by comparing their execution times against the following programs:
\begin{packeditemize}   
    \item \textbf{DIQL}: The algorithms have been expressed using DIQL \cite{FegarasN18} queries. In particular, the fixpoint operation is expressed in terms of the more generic \opname{repeat} operator of the DIQL language. 
    We have written the queries in such a way that they compute the fixpoint more efficiently using the algorithm mentioned in ~\ref{sec:locexec}. The DIQL system is available at~\cite{diql-github}.
    \item \textbf{Emma}: We used the example provided by Emma authors \cite{emma-github} to compute the TC queries, and we wrote modified versions to compute the SP and the path planning examples. The Emma system is available at~\cite{emma-github}.
    \item \textbf{\mumonoidnoPA}: \mumonoid{} without the application of $\pushingaggregRule$ to assess the impact of the $\pushingaggregRule$ rule. 
    \item \textbf{\mumonoidnoPdist{}}: \mumonoid{} without the application of $\fixpointdistRule$ to assess the impact of the $\fixpointdistRule$ rule.
\end{packeditemize}
All these systems run on the same experimental setup presented above and on the same Spark plateform.

We have also written the DIQL queries in such a way they apply $\pushingfilterRule$. Such a pre-filtering was not possible for Emma because the programs perform a non linear fixpoint. Trying to write a linear version leads to an exception in the execution. We were not able to write an Emma program that computes movie recommendations. Iterating over a users own movies leads to an exception.

\paragraph{Datasets.}
We use two kinds of datasets:
\begin{packeditemize}
    \item Real world graphs of different sizes, presented in Table~\ref{tab:graphsize}, including a knowledge graph (the Yago \cite{yago} dataset\footnote{\label{footnote:yago}We use a cleaned version of the real world dataset Yago 2s \cite{yago}, that we have preprocessed in order to remove duplicate RDF \cite{rdf} triples (of the form <source, label, target>) and keep only triples with existing and valid identifiers. After preprocessing, we obtain a table of Yago facts with 83 predicates and 62,643,951 rows (graph edges).
    
    For this dataset, transitive closures are computed for the \texttt{isLocatedIn} edge label.
    }), a social network graph (Facebook), and a scientific collaborations network (DBLP) taken from \cite{snap}.
    \item Synthetic graphs shown in Table~\ref{tab:graphsize}, generated using the Erdos Renyi algorithm that, given an integer $n$ and a probability $p$, generates a graph of $n$ vertices in which two vertices are connected by an edge with a probability $p$. \rndfile{n}{p} denotes such a synthetic graph, whereas \rndfile{n}{p}\_W denotes a \rndfile{n}{p} graph with edges weighted randomly (between $0$ and $5$).

    Other synthetic graphs are:
 \begin{packeditemize}
    \item{\rndfile[flight]{n}{p}}: where edges are taken from \rndfile{n}{p} with random depart and arrival times and duration assigned to them.
    \item{\rndfile[c]{n}{p}}: serialized object RDD files representing paths between cities. It is also generated from \rndfile{n}{p}, each city has been assigned up to 10 random landmarks.
    \item{u\_n}: serialized object RDD files of $n$ users, each assigned up to 15 random movies.
 \end{packeditemize}

\end{packeditemize}
\begin{table}
    \begin{small}
    \begin{minipage}[t]{8cm}
        \bgroup
        \def\arraystretch{1.1}%
        \begin{tabular}{|r|r|r|r|r}\hline
        Dataset & Edges & Nodes & TC size  \\ \hline \hline
        \rndfile{10k}{0.001} & 50,119 & 10,000 & 5,718,306\\
        \rndfile{20k}{0.001} & 199,871 & 20,000 & 81,732,096  \\
        \rndfile{30k}{0.001} & 450,904 & 30,000 & 255,097,974\\        
        \rndfile{10k}{0.005} & 249,791 & 10,000 & 39,113,982 \\
        \rndfile{10k}{0.01} & 499,486 & 10,000 & 45,098,336 \\
        \rndfile{40k}{0.001} & 799,961 & 40,000 & 531,677,274 \\
        \rndfile{50k}{0.001} & 1,250,922 & 50,000 & 906,630,823   \\ \hline
        \end{tabular}
        \egroup
    \end{minipage}
    \begin{minipage}[t]{5cm}
        \bgroup
        \def\arraystretch{1.1}%
        \begin{tabular}{|r|r|r|r}\hline
        Dataset & Edges & Nodes  \\ \hline\hline
        Yago  & 62,643,951 & 42,832,856 \\ \hline
        Facebook & 88,234 & 4,039 \\
        DBLP & 1,049,866 & 317,080     \\      \hline         
        \end{tabular}
        \egroup
    \end{minipage}
    \caption{Synthetic and real graphs used in experiments.}\label{tab:graphsize} 
    \end{small}
\end{table}

\begin{figure*}[h]
    \centering
    \includegraphics[width=0.8\textwidth]{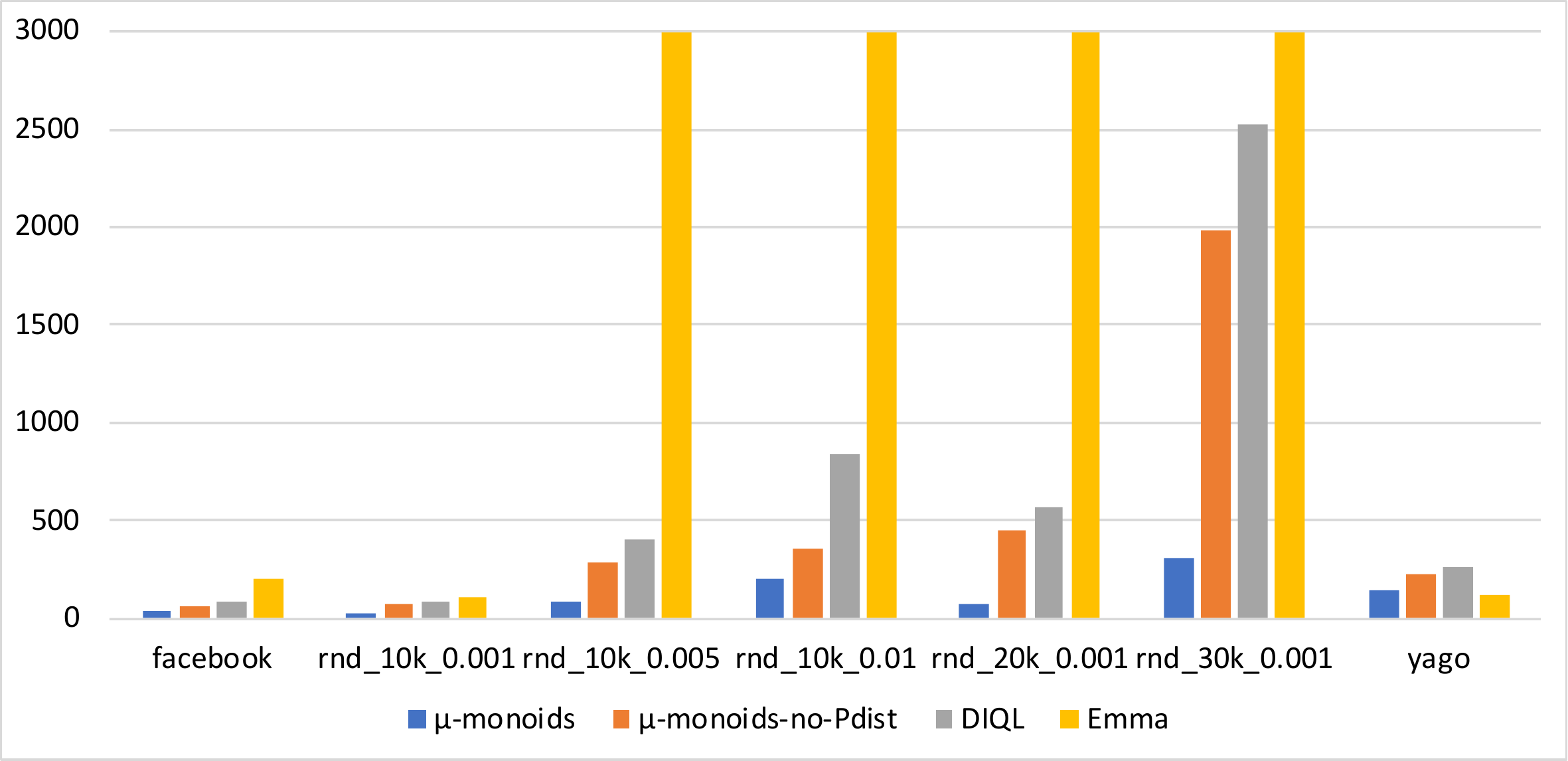}
    \captionof{figure}{\label{fig:tc}
     TC running times.}
\end{figure*}

\begin{figure*}[h]
    \centering
    \includegraphics[width=0.8\textwidth]{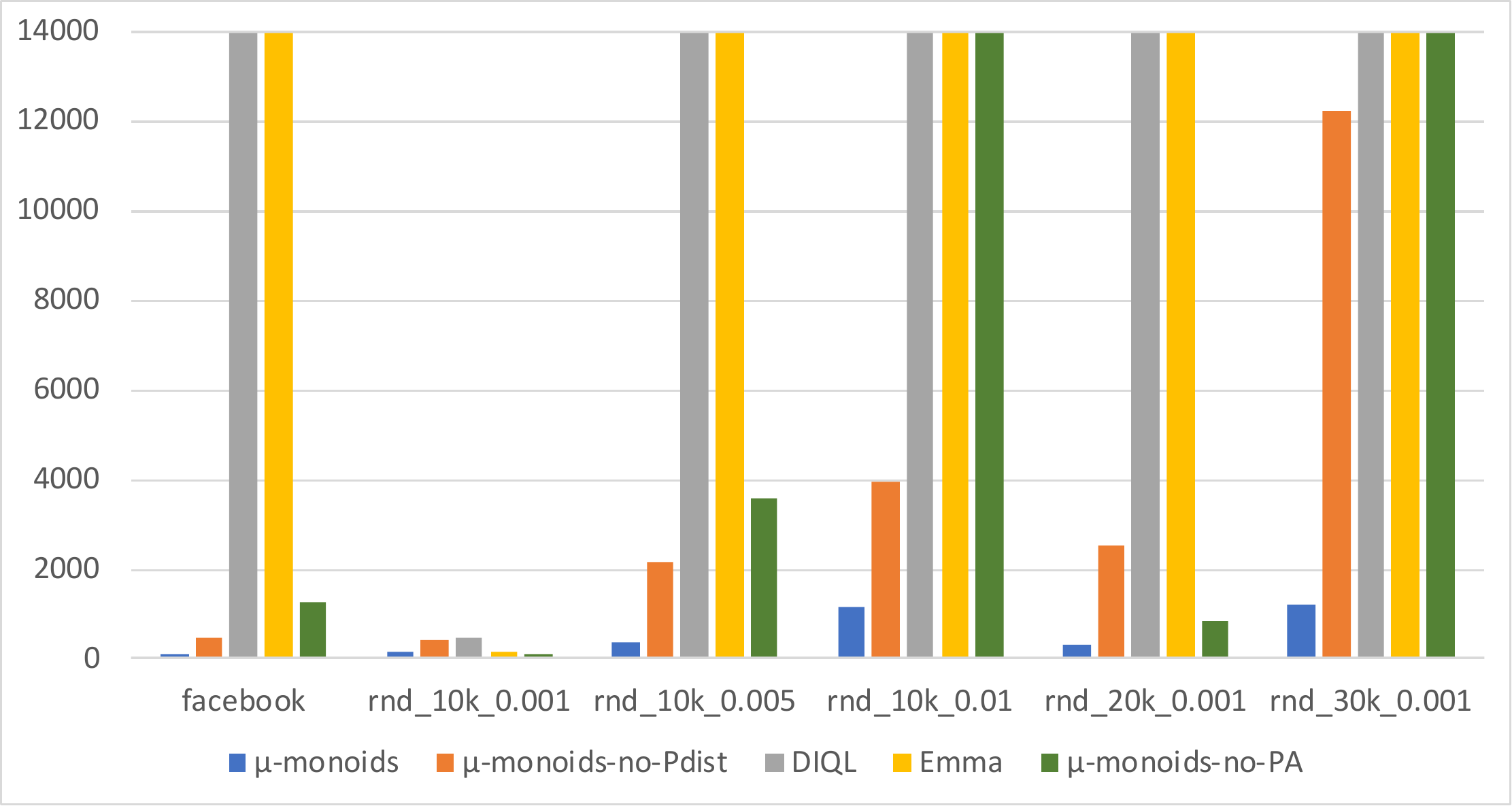}
    \captionof{figure}{\label{fig:sp}
    SP running times.}
\end{figure*}

\begin{figure*}[h]
    \centering
    \includegraphics[width=1\textwidth]{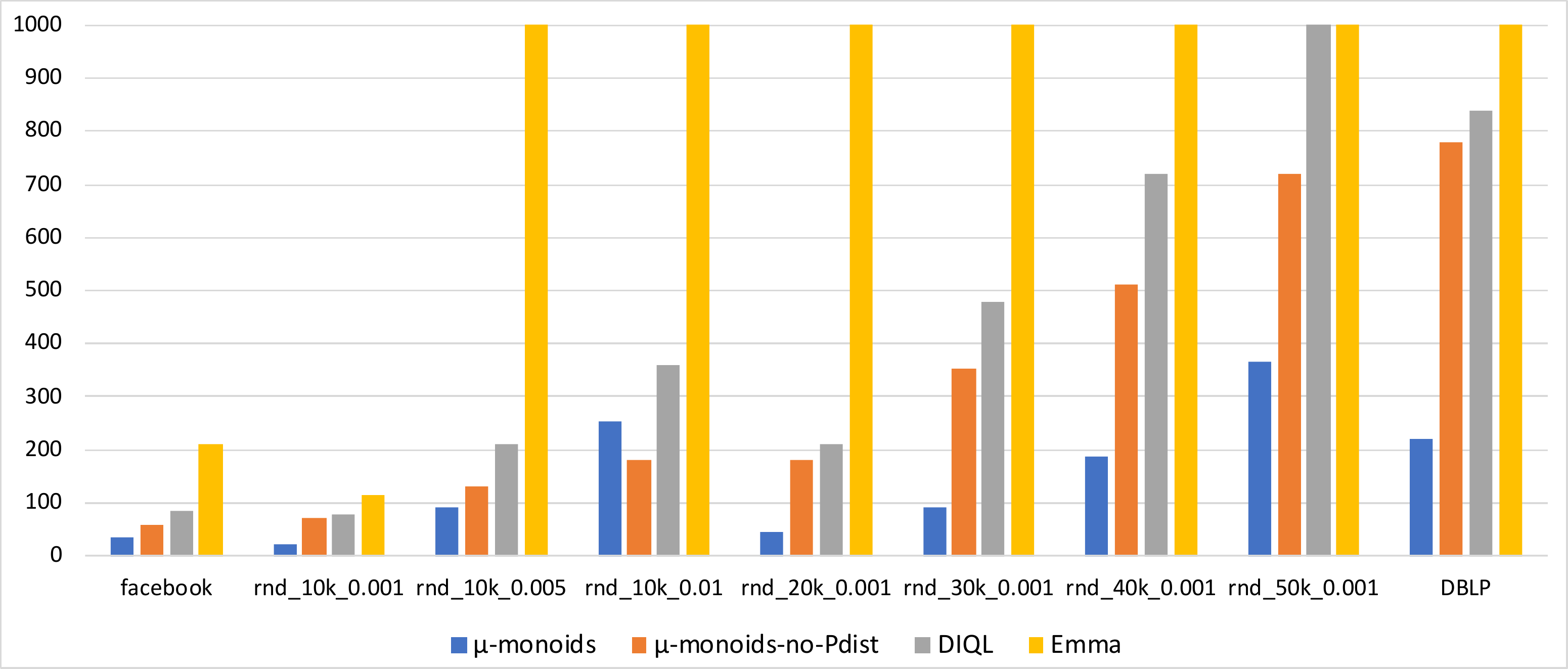}
    \captionof{figure}{\label{fig:tcfilter}
     TC filter running times.}
\end{figure*}
\begin{figure*}[h]
    \centering
    \includegraphics[width=1\textwidth]{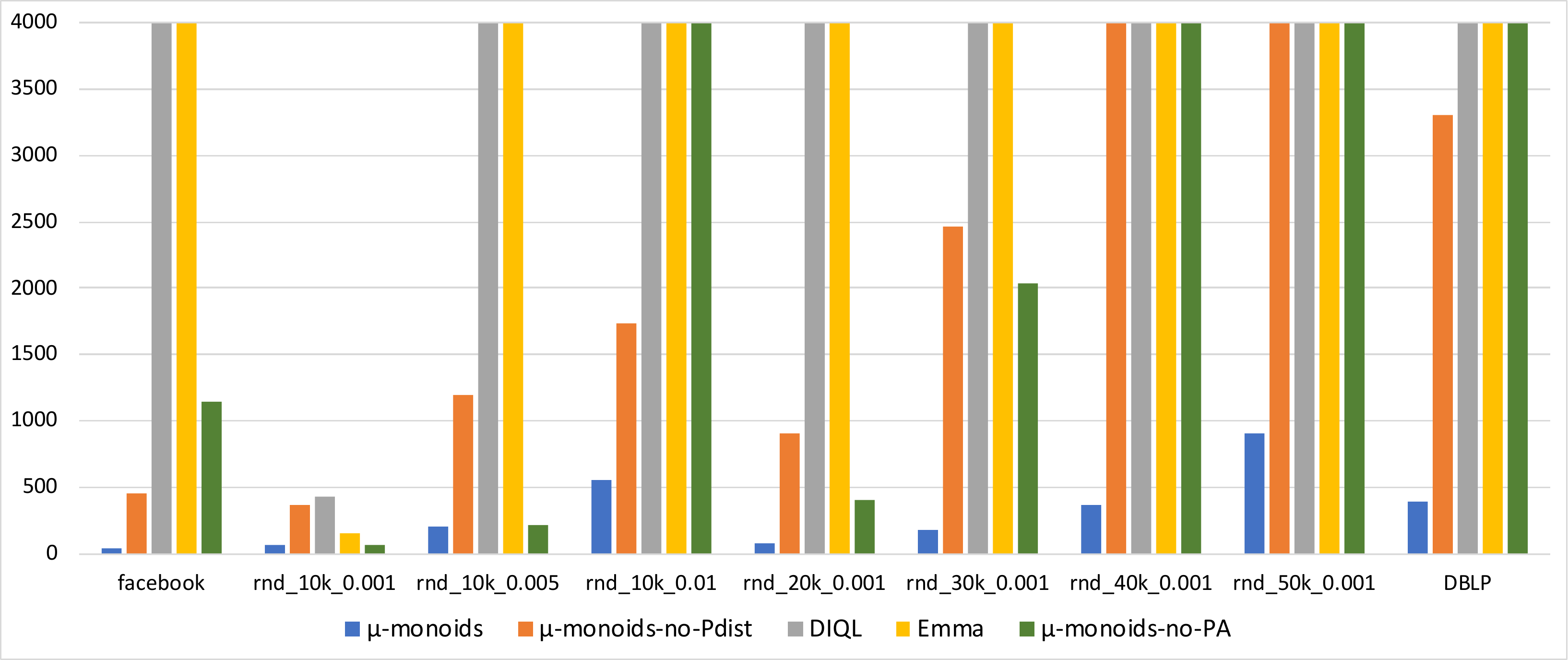}
    \captionof{figure}{\label{fig:spfilter}
    SP filter running times.}
\end{figure*}

\begin{figure*}[h]
    \centering
    \includegraphics[width=0.8\textwidth]{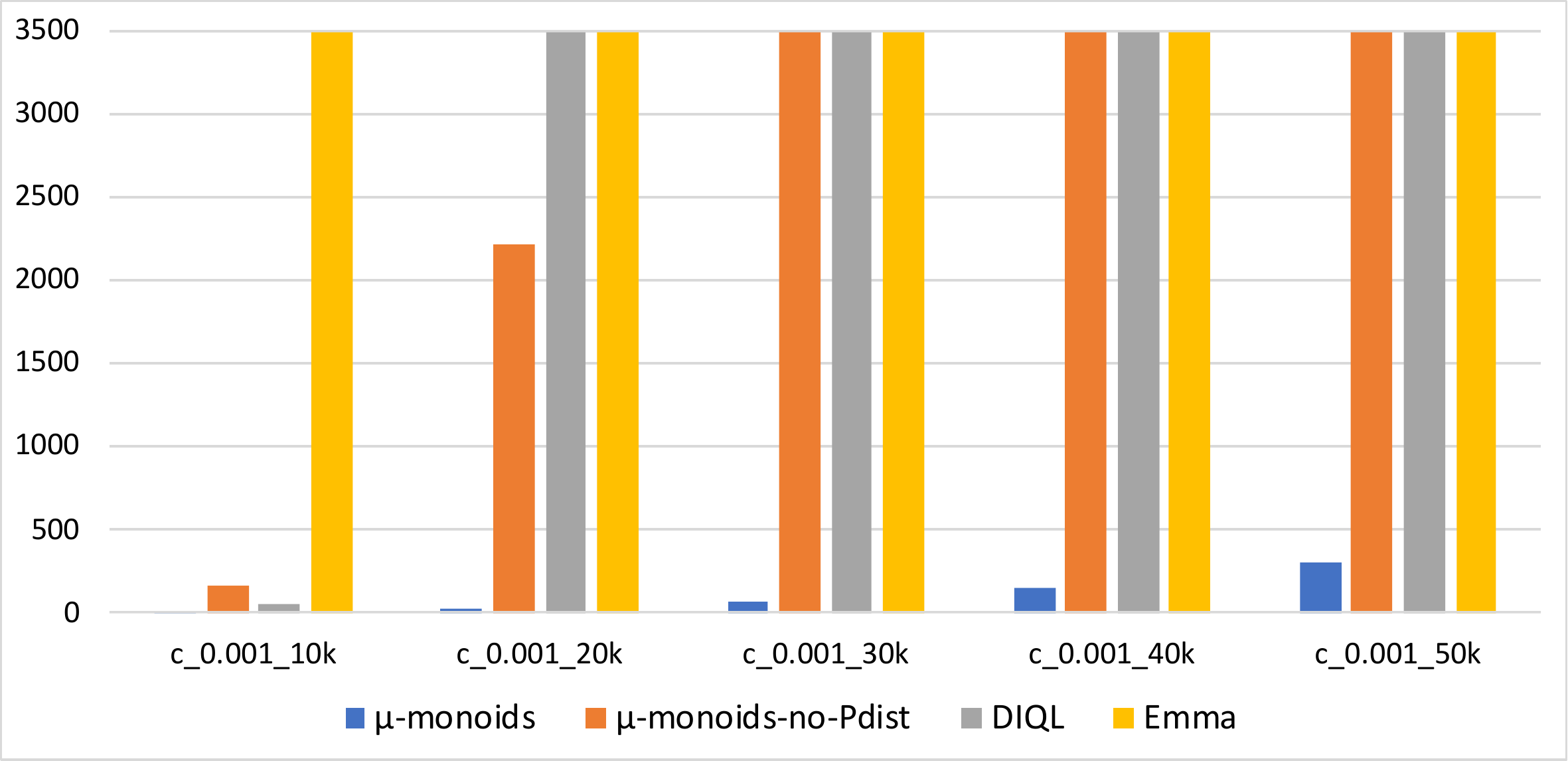}\;
    \captionof{figure}{\label{fig:pathplannign}
     Path planning running times.}
\end{figure*}

\begin{figure*}[h]
    \makebox[\textwidth][c]{
    \includegraphics[width=0.743\textwidth]{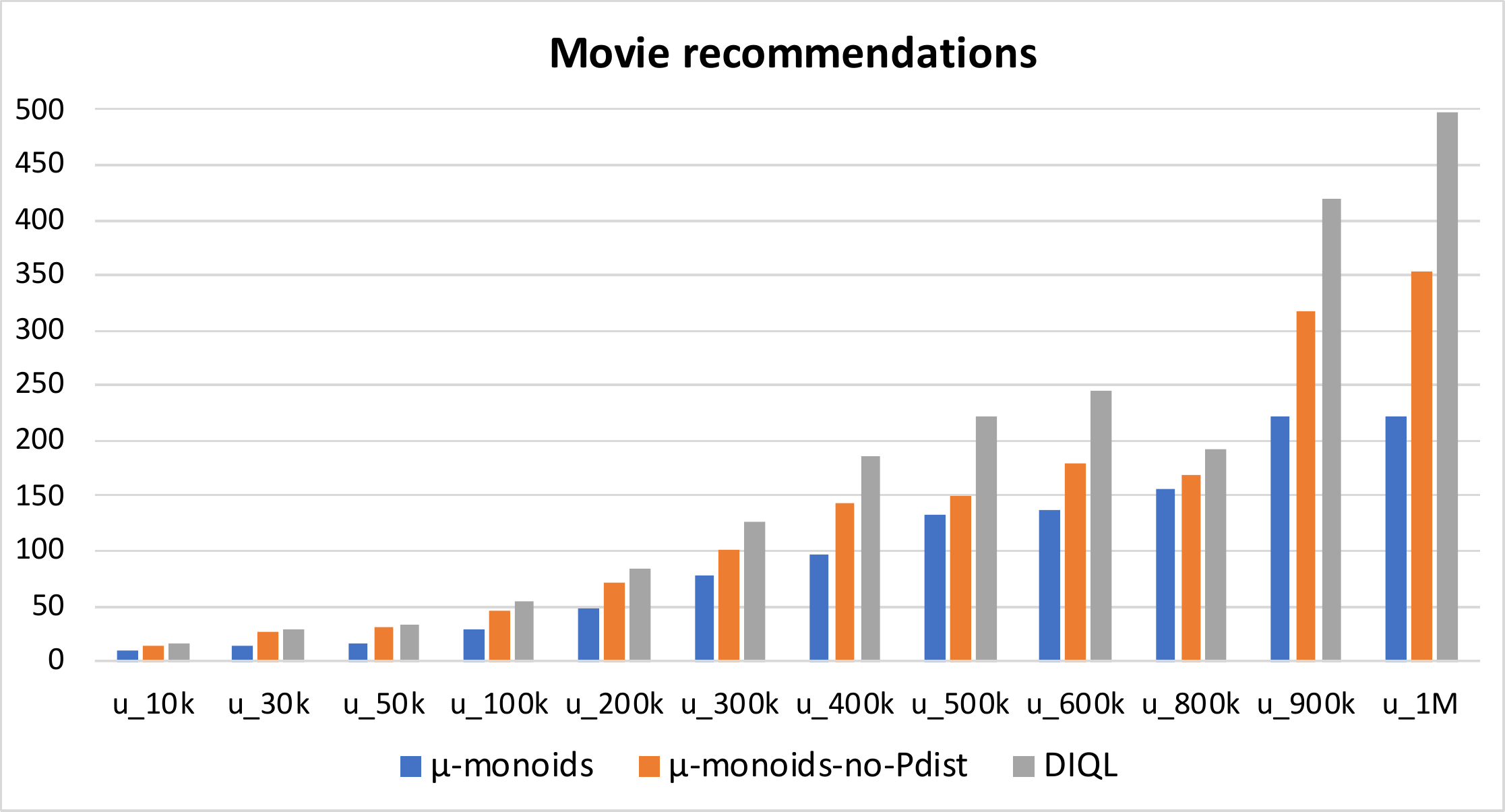}\;
    \includegraphics[width=0.29\textwidth]{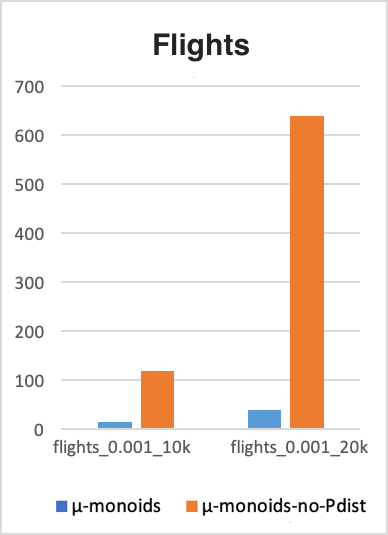}
    }
    \captionof{figure}{\label{fig:movierecflights}
    Movie recommendations and Flights running times.}
\end{figure*}

\paragraph{Results summary.}

Figures~\ref{fig:tc}~to~\ref{fig:movierecflights} show the comparison of the programs running times (reported in seconds on the y-axes). A bar reaching the maximal value on the y-axis indicates a timeout. Each data point represents the average of 5 runs. 

We observe that the programs generated by \mumonoid{} almost systematically 
outperform the other program versions and always outperform the DIQL and Emma systems.

Comparison between \mumonoid{} and \mumonoidnoPdist{} shows the impact of the $\fixpointdistRule{}$ rule. 
It can be noticed that the speedup achieved by this rule increases with the data size.
The only case where \mumonoid{} is slower than \mumonoidnoPdist{} is on the $\rndfile{10k}{0.01}$ dataset in Fig.~\ref{fig:tcfilter}.
This is due to the graph topology as this dataset is the densest graph tested. This means that the size of the intermediate results generated during the fixpoint computation is large, 
which puts more strain on the parallel tasks. The fixed number of parallel tasks (Sec.~\ref{sec:ruleseffects}) is too small. We observe in practice that, in this case, increasing the number of partitions improves the performance of the system.

In the SP, SP filter and path planning programs, both $\fixpointdistRule{}$ and $\pushingaggregRule$ are applied in \mumonoid{}. We notice that the speedup of \mumonoid{} in comparison to DIQL and Emma is even more important in these cases.
Comparison between \mumonoid{} and \mumonoidnoPA{} shows the impact of the $\pushingaggregRule$ rule alone.
It can also be observed that applying $\pushingaggregRule$ and not $\fixpointdistRule$ (\mumonoidnoPdist{}) can be faster than applying $\fixpointdistRule$ and not $\pushingaggregRule$ (\mumonoidnoPA{}) and the other way round depending on the cases. It is the combination of these two rules that leads to the best performances. 

This experimental comparison shows the benefit of the plan that distributes the fixpoint.
It also highlights the benefits of the approach that synthesises code: generating programs that are not natural
 for a programmer to write, like the distributed loop to compute the fixpoint.

\newpage
\packedsection{Related works} \label{sec:relatedworks}

The idea of using an intermediate representation (or an algebra) for representing user queries and performing automatic optimization originates from the work of Codd~\cite{codd1970relational}. He proposed the idea of a separation between the internal representation (physical storage) from the logical representation of data. The idea is to offer a level of abstraction to represent data and operate upon it via a universal language which is independent from implementation details and possible changes to how data is physically stored and retrieved. This insight led to the relational algebra being widely adopted by database systems and extensively studied in database research. It also led to the standard SQL language. 
SQL is a \emph{Domain Specific Language (DSL)} that is called from within a general purpose language (also called \emph{the host language}). It gets translated to a relational algebra term that gets optimized then translated to a physical execution plan.
Work in~\cite{acmsurvey2021} surveys state of the art approaches for handling iterations in distributed systems and classifies them to different categories. Among these categories are the relational algebra approach mentioned earlier, and the functional approach that offer higher-order functions for specifying control flow (such as loops). The present work belongs to this latter category.
 In this section we review related literature along two lines of work: works that are based on the relational model, and works that are based on more generic data models.
 
\subsection{Works based on the relational formalism}
The relational model is based on n-ary relations to represent entities and relationships between them. 
An n-ary relation is a set of rows. Each row consists of a tuple of n records of atomic types.
To operate on these relations, \cite{codd1970relational} proposed the relational algebra. Relational algebra offers operations on relations such as projection, selection, and join, as well as a number of rewrite rules that aim to optimize expressions regardless of their initial shape. 

Another prevalent formalism based on the relational model is Datalog~\cite{datalogsurvey2018}. A Datalog program is composed of rules that infer new facts from previously known facts. Facts are expressed as predicates depending on a fixed number of variables. They can thus be seen as relations. Optimization techniques such as Magic Sets~\cite{magic_set1, magic_set2} or Demand transformation~\cite{tekle2011more} are proposed to optimize Datalog programs.

Regarding the ability to express recursion, a number of formalisms that extend RA with a recursive operator have been proposed~\cite{agrawal_alpha:_1988, Aho79}. The algebra proposed in \cite{mura-sigmod20} provides more optimizations of recursion than previous works on RA and recursive Datalog. However, it is limited to the centralized setting and to relational algebra. 
Regarding distribution,
the Spark SQL~\cite{sparksql2015} library enables the user to write SQL queries and process relational data using \texttt{dataset}s or \texttt{dataframe}s. Queries are optimized using the Spark Catalyst engine and executed in a distributed way on the Spark platform.
BigDatalog~\cite{bigdatalog} is a system that studies the distribution of Datalog programs on Spark. 
Compared to previous distributed Datalog systems such as Socialite~\cite{socialite2013} and Myria~\cite{myria2015}, it achieves better performances.
The BigDatalog system
uses the Datalog GPS technique~\cite{seib1991} that analyses Datalog rules to identify decomposable Datalog programs and determine how to distribute data and computations. These ideas are tied to Datalog and are not applicable to other formalisms. 
In contrast, the present work proposes a new distribution method designed for a more generic algebra.

\subsection{Works based on more generic formalisms}
In the relational model, data consists of relations that are sets of tuples of atomic values.
A more generic data model are collections of arbitrary homogeneous types.
In Sec.~\ref{sec:collmon} we discussed the 3 types of data structures which we call \emph{collections}: lists, bags, and sets.
They are, along with other data structures, part of what is called the Boom hierarchy of types~\cite{Bunkenburg1993}.
The author of this paper presents data structures of that hierarchy as free algebras. A data structure value can either be empty $[~]$, a singleton $[a]$, or a combination of two values using a binary operator $c1 \lunion c2$.
$\lunion$ can obey to a combination of four algebraic laws: unit (it has a unit element), associativity, commutativity, and idempotence.
Different combinations of laws lead to different types of structures. \cite{Bunkenburg1993} defines 16 types of data structures for all possible combinations of these laws. For instance, \emph{tree} is a data structure where only the first law is satisfied: it is not associative (and therefore not a monoid). Lists are the data structure from the hierarchy obeying only unit and associativity. Bags are obtained when we add commutativity, and sets when we add idempotence. These three structures, which are monoids, are called \emph{collection monoids} in the work of~\cite{fegaras-jfp2017}. It is on this basic notion this work builds the \emph{monoid algebra}.
Collections can also be seen as a particular case of Algebraic Data Types which constitutes the basic notion of the Emma language approach~\cite{emma}. Both of these approaches propose an algebra for distributed collections. We review these works in more detail below.

It was shown in \cite{BunemanNTW95} that monads can be used to generalize nested relational algebra to different collection types and complex data.
Wadler also explored and developed monad comprehensions \cite{Wadler92} and ringad comprehensions \cite{Gibbons16}, inspired by the early works on list comprehensions \cite{Turner2016,PeytonJones:1987:IFP:1096899}. 
These ideas were used in LINQ \cite{MeijerBB06}, Ferry~\cite{grust-sigmod09}, and Emma \cite{alexandrovKKM16} which are comprehensions-based programming languages. 
To be evaluated, LINQ and Ferry queries get translated into an intermediate form that can be executed on relational database systems supporting SQL. As they target relational database systems, the set of host language expressions that can be used in query clauses like selection and projection is restricted. In addition, they do not analyse comprehensions to make optimizations. 
 Emma \cite{alexandrovKKM16} is a comprehensions-based language similar in spirit, but which rather targets JVM-based parallel dataflow engines (such as Spark and Flink). DryadLinq~\cite{dryadlink2008} proposes a system that distributes LINQ queries on the Dryad platform~\cite{dryad2007}. In these works recursion is expressed using host language loops that are not modelled in the algebra, hence with no optimisation provided.
 The idea of using monoids and monoid homomorphisms for modeling computations with data collections originates from the works found in \cite{TannenBN91,TannenBO91}.
Fegaras proposed a monoid comprehension calculus \cite{FegarasM00} which later evolved in the monoid algebra presented in \cite{fegaras-jfp2017}. It proposes an algebra based on monoid homomorphisms therefore with parallelism at its core: a homomorphic operation $H$ on a collection is defined as the application of $H$ on each subpart of the collection, results are then gathered using an associative operator. 
Distributed collections are modelled using the union representation of bags, and collection elements can be of any type defined in the host language. The algebra allows for defining second order operators such as \opname{flatmap} and \opname{reduce} that can take a UDF written in the host language as an argument. The present work further builds on this approach and proposes a generic criteria (Sec.~\ref{sec:pushingfilter}) using the host language type checking system that examines those UDFs in order to determine whether the $\pushingfilterRule$ optimization can take place.
The monoid algebra~\cite{fegaras-jfp2017} has a \texttt{repeat} operator, however no optimization is provided for this operator.
The authors of~\cite{FegarasN18} designed DIQL (a DSL that translates to the monoid algebra). 
Using reflection of the host language (Scala in this work) and quotations, queries of this DSL can be compiled and type checked seamlessly with the rest of the host language code. In fact, Emma uses the same approach as well.
This approach offers more optimization opportunities than approaches like the Spark and Flink API. As argued in~\cite{emma}, even though these APIs offer a DSL that is well integrated with the host language and allow for expressing general purpose computations, they suffer from the difficulty of automatically optimizing programs. This is due to the limited program context available in the intermediate representation of the DSLs. For instance, arguments to second order operations are treated as black box functions which means that they cannot be analyzed and transformed to make automatic optimizations.

\packedsection{Conclusion} \label{sec:conclu}

We propose to extend the monoid algebra with a fixpoint operator that models recursion. The extended \mumonoid{} algebra is suitable for modeling recursive computations with distributed data collections such as the ones found in big data frameworks. The major interest of the introduced ``$\mu$'' fixpoint operator is that it can be considered as a monoid homomorphism and thus can be evaluated by parallel loops with one final merge rather than by a global loop requiring network overhead after each iteration.

We also propose rewriting rules for optimizing fixpoint terms: we show when and how filters can be pushed into fixpoints. In particular, we find a sufficient condition on the repeatedly evaluated term ($\varphi$) regardless of its shape, and we present a method using polymorphic types and a type system such as Scala's to check whether this condition holds. We also propose a rule to prefilter a fixpoint before a join. The third rule allows for pushing aggregation functions inside a fixpoint.

Experiments suggest that: (i)  Spark programs generated by the systematic application of these optimizations can be radically different from -- and less intuitive -- than the input ones written by the programmer;  (ii) generated programs can be significantly more efficient.  This illustrates the interest of developing optimizing compilers for programming with big data frameworks.

\newpage
\bibliographystyle{alphaurl}
\bibliography{paper}
\end{document}